\documentclass[twocolumn,aps,prc,superscriptaddress,floatfix,nofootinbib]{revtex4-2}
\usepackage{subcaption}
\usepackage{bm}
\usepackage{float}
\usepackage{graphicx}
\usepackage{booktabs}
\usepackage{tabularx}
\usepackage{amsmath}
\usepackage{color}
\usepackage{comment}
\usepackage[colorlinks,citecolor=blue,urlcolor=blue,linkcolor=blue]{hyperref}
\usepackage{newtxtext,newtxmath,booktabs,siunitx}
\usepackage{dcolumn}% Align table columns on decimal point
\usepackage{booktabs}
\usepackage{multirow}
\usepackage{ulem}

\renewcommand{\d}{\mathrm{d}}
\newcommand{\kt}{\bm{k}_T}
\newcommand{\pt}{\bm{p}_T}

\newcommand{\blue}[1]{{\color{blue}{#1}}}

\newcommand{\red}[1]{{\color{red}{#1}}}
\newcommand{\Remove}[1]{{\color[rgb]{0.8,0.3,0.3}\ifmmode\text{\sout{$#1$}}\else\sout{#1}\fi}}

\newcommand{\com}[1]{\textcolor{red}{#1}}

\begin{document}

\title{%Hadron production in forward rapidity region in pp collisions at LHC energies \\ from an event generator based on color glass condensate
%\\
%$\to$\\
Forward hadron production in pp collisions at LHC energies \\ from an event generator based on the color glass condensate framework
}

\author{Hirotsugu Fujii}
\affiliation{Nishogakusha University,  Tokyo 102-8336, Japan}

\author{Tetsufumi Hirano}
\affiliation{Department of Physics, Sophia University, Tokyo 102-8554, Japan}

\author{Kazunori Itakura}
\affiliation{Nagasaki Institute of Applied Science, Nagasaki 851-0193, Japan}

\author{Yasushi Nara}
\affiliation{Akita International University, Akita 010-1292, Japan}

\author{Shujun Zhao}
\email{zhaosj@sophia.ac.jp}
\affiliation{Department of Physics, Sophia University, Tokyo 102-8554, Japan}

\begin{abstract}
  We investigate inclusive forward single-hadron production in high-energy proton--proton collisions using a CGC-inspired Monte Carlo event generator, MC-CGC. We carried out a systematic study of the sensitivity of the running-coupling Balitsky-Kovchegov (rcBK) evolution equation to its initial conditions by comparing three parameterizations: the McLerran-Venugopalan (MV) model and its two HERA DIS-constrained variants, MV$^\gamma$ and MV$^e$. Our results indicate that the current LHCb data favor the MV$^\gamma$ and MV$^e$ models, while the differences from the original MV model become more pronounced at higher transverse momentum and at mid-rapidity. As a complementary analysis, we also compared the dilute-dense (DHJ factorization) and dense-dense ($k_T$ factorization) frameworks. We found that the $k_T$ factorization framework provides a better description of the particle production spectra at mid-rapidity than the DHJ framework, where both the projectile and target are in the dense regime at LHC energies. Predictions for the FoCal measurements at ALICE, including the production of identified neutral mesons and jets, are also presented. 
\end{abstract}
\maketitle

\section{Introduction}

%Understanding Quantum Chromodynamics (QCD) dynamics in high-energy collisions from the first principles has long been a central challenge in strong interaction physics. 
A first-principles understanding of the high-energy limit of Quantum Chromodynamics (QCD) remains an important challenge in strong interaction physics. In the high-energy limit, particle production is dominated by small Bjorken-$x$ gluons, whose densities grow rapidly, leading to non-linear dynamics and the emergence of a universal saturation regime of QCD. To systematically describe this regime,  the Color Glass Condensate (CGC) framework \cite{Iancu:2003xm, McLerran:1993ni, McLerran:1993ka, Gelis:2010nm} has been developed over the past few decades as an effective field theory of small-$x$ gluons, providing a consistent theoretical tool for both analytical calculations and phenomenological applications.

Evidence supporting the saturation picture has emerged from deep inelastic scattering (DIS) at HERA through geometric scaling with the saturation momentum $Q_s$ \cite{Golec-Biernat:1998zce, Stasto:2000er}, as well as from measurements in proton (deuteron)–nucleus collisions at Relativistic Heavy Ion Collider (RHIC) and the Large Hadron Collider (LHC), including forward particle suppression~\cite{BRAHMS:2004xry, STAR:2006dgg}, reduction of back-to-back correlations~\cite{PHENIX:2011puq, STAR:2021fgw}, and emergence of long-range rapidity correlations (“ridge”) \cite{Albacete:2014fwa, CMS:2012qk, ATLAS:2014qaj, ALICE:2012eyl}. However, these observations are not yet sufficient to establish parton saturation and the CGC picture unambiguously, as alternative QCD mechanisms may also account for them~\cite{Accardi:2004ut, Armesto:2006ph, Arnaldi:2009it}.

The CGC framework is also essential for characterizing the initial conditions of quark--gluon plasma (QGP) formation in high-energy heavy-ion collisions, where the collision of two saturated gluon fields produces a glasma that subsequently evolves through non-equilibrium dynamics towards a QGP \cite{Lappi:2006fp, Berges:2020fwq, Gelis:2012ri}. We note that, in small systems, initial-state parton correlations associated with saturation dynamics can give rise to flow-like azimuthal structures, which have contributed to ongoing discussions on the origin of collective behavior in small collision systems \cite{Schlichting:2016sqo, Grosse-Oetringhaus:2024bwr}. 

To further clarify the situation, parton saturation is being experimentally explored at the LHC~\cite{LHCb:2021vww, LHCb:2022tjh, LHCb:2022dmh, LHCb:2022ahs}, with future measurements such as the ALICE forward calorimeter (FoCal) \cite{ALICE:2020mso} expected to further extend the accessible kinematic range. It is also one of the primary objectives of the planned Electron–Ion Collider (EIC) \cite{AbdulKhalek:2021gbh} and plays an important role in the interpretation of ultra-high-energy cosmic ray interactions \cite{Engel:2011zzb}. 

The phenomenological applications of the CGC framework have made significant progress, driven by the successful global analyses \cite{Albacete:2009fh, Albacete:2010sy, Dai:2026nzp} of the DIS data at small $x$ based on the running-coupling Balitsky--Kovchegov (rcBK) equation \cite{Balitsky:2006wa, Kovchegov:2006vj} with appropriate initial conditions (see also \cite{Beuf:2020dxl}). These analyses provide a quantitative constraint on the dipole scattering amplitude in the small-$x$ regime. Building on these developments, particle production in dilute--dense collisions has been extensively studied within the so-called hybrid formalism, also known as the Dumitru--Hayashigaki--Jalilian-Marian (DHJ) approach \cite{Dumitru:2005gt}. In this framework, the dense target is described in terms of the dipole amplitudes, which enables a unified description of inclusive single hadron production at forward rapidities in proton--proton and proton--nucleus collisions \cite{Albacete:2010bs, Fujii:2011fh, Albacete:2012xq, Lappi:2013zma}. 
In dense-dense collisions, single-particle production is often modeled in $k_T$-factorized expressions using the unintegrated gluon distributions. This approach can incorporate saturation effects phenomenologically~\cite{Kharzeev:2004if}, although $k_T$ factorization is known to break down due to multiple interactions between the two dense sources~\cite{Blaizot:2004wu, Gelis:2008rw}.

%\red{Studies on inclusive single-particle production extended beyond the leading order (LO), with next-to-leading order (NLO) corrections aimed at improving theoretical precision~\cite{Balitsky:2007feb, Chirilli:2011km, Shi:2021hwx, Liu:2022ijp}. However, this NLO extension has involved a complicated history. The first numerical evaluations found that the NLO cross-section became unphysically negative at large transverse momentum ($p_T$)~\cite{Stasto:2014sea}, while around the same time the NLO BK equation was found to be numerically unstable~\cite{Lappi:2015fma}. A consistent treatment of rapidity factorization requires the use of NLO expressions for both the hard impact factor and the evolution equation~\cite{Chirilli:2011km}. Although implementing kinematical constraints successfully resolved the instabilities in the NLO BK equation~\cite{Iancu:2015vea}, and enabled stable NLO calculations for DIS~\cite{Beuf:2020dxl}, the single-particle cross-section still exhibited unphysically negative values at high $p_T$ and had a large dependence on the prescription for the QCD running coupling~\cite{Ducloue:2017mpb}. Since around 2021, improvements in accounting for large threshold corrections have successfully restored positive cross-sections~\cite{Shi:2021hwx, Liu:2022ijp}.}

Studies on inclusive single-particle production have been extended further beyond the leading order (LO), with next-to-leading order (NLO) corrections aimed at improving theoretical accuracy~\cite{Balitsky:2007feb, Chirilli:2011km, Shi:2021hwx, Liu:2022ijp}. However, this NLO extension has involved a complicated history. The first numerical evaluations found that the NLO cross-section became unphysically negative at large transverse momentum ($p_T$) ~\cite{Stasto:2014sea}, while around the same time the NLO BK equation was found to be numerically unstable~\cite{Lappi:2015fma}. A consistent treatment of rapidity factorization requires the use of NLO expressions for both the hard impact factor and the evolution equation~\cite{Chirilli:2011km}. Although implementing kinematical constraints successfully resolved the instabilities in the NLO BK equation~\cite{Beuf:2014uia, Iancu:2015vea} and enabled stable NLO calculations for DIS~\cite{Beuf:2020dxl}, the single-particle cross-section still exhibited unphysically negative values at high $p_T$ and had a large dependence on the prescription for the QCD running coupling~\cite{Ducloue:2017mpb}. Since around 2021, improvements in accounting for large threshold corrections have successfully restored positive cross-sections~\cite{Shi:2021hwx, Liu:2022ijp}.

Multi-particle correlation observables~\cite{Albacete:2010pg, Marquet:2007vb, Baier:2005dv, Jalilian-Marian:2004vhw, Dumitru:2008wn, Gelis:2008rw, Lappi:2012nh, Caucal:2025zkl} have also been explored recently via higher-point correlation functions. The description of these observables requires the evaluation of
multi-point Wilson-line correlators (such as quadrupoles), whose rapidity evolution goes beyond the dipole approximation and is governed by the JIMWLK\footnote{Jalilian-Marian--Iancu--McLerran--Weigert--Leonidov--Kovner.} equation~\cite{Jalilian-Marian:1997jhx, Iancu:2001ad}. Systematic studies of multi-gluon production and multi-particle cumulants ~\cite{Agostini:2021xca, Agostini:2022ctk, Schenke:2019pmk, Kohara:2023qnp} have been performed within the JIMWLK framework of the CGC.

Although these developments constitute important steps toward a more complete understanding of saturation dynamics, the increasing complexity of the formalism presents substantial challenges for practical applications. The treatment of the other effects, including the contributions from soft non-perturbative regions or the beam remnants, also needs to be included consistently. In this context, an event-by-event approach incorporating the saturation physics can provide a useful and flexible tool to bridge the gap between theoretical developments and phenomenological studies. Some investigations on the multiplicity distribution~\cite{Schenke:2012hg}, anisotropic flow~\cite{Schenke:2016lrs}, and charm production~\cite{Bhattacharya:2023zei} have been made, while a full event-generator still requires further development.

In the previous work, we adopted a complementary strategy and focused on constructing an event generator based on the leading-order CGC framework \cite{Deng:2014vda}. By incorporating the essential features of saturation physics at LO into a Monte Carlo framework, this approach allows us to investigate forward particle production in terms of fully exclusive final states, while maintaining a level of theoretical control and numerical stability suitable for phenomenological applications. In this work, we update our previous CGC-based event generator model~\cite{Deng:2014vda} to achieve a consistent description of recent data from experiments and to provide predictions for future measurements. We will also investigate observables that were not discussed in the previous work. Given the theoretical complexities, implementing a fully NLO-consistent treatment, including NLO rcBK evolution and the NLO hybrid formalism, in practical event generators remains highly challenging at present. The present work therefore employs the running-coupling BK (rcBK) equation at LO and does not incorporate the aforementioned kinematical improvements or NLO treatment in the evolution or in the single-inclusive production cross section.
%\red{In this work, we update our previous CGC-based event generator model \cite{Deng:2014vda} to achieve a consistent description of recent data from the LHCb experiment and to provide predictions for future measurements. We will also investigate observables that were not discussed in the previous work. Given the theoretical complexities, implementing a fully NLO-consistent treatment, including NLO rcBK evolution and the NLO hybrid formalism, in practical event generators remains highly challenging at present. The present work therefore employs the running-coupling BK (rcBK) equation at LO and does not incorporate the aforementioned kinetics improvements or NLO treatment in the evolution or in the single-inclusive production cross section.}

This paper is organized as follows. In Sec.~\ref{sec:mc-cgc}, we explain our updated model 
implementing the DHJ formula and the $k_T$ factorization formula within the PYTHIA framework in some detail.
In Sec.~\ref{sec:Evaluation-of-Existing-Measurements}, we compare our model with forward inclusive single-particle production measured at the LHC and RHIC, and assess its overall quality. We also confront our model with the mid-rapidity LHC data, showing that the $k_T$-factorization approach provides a better description than the DHJ model. In Sec.~\ref{sec:predicsion-for-future-focal-measurements}, we present predictions for the FoCal measurement, including several bulk observables of identified hadrons at more forward rapidities. We also show results for jet observables in this kinematic region. Finally, Sec.~\ref{sec:conclusion} is devoted to a summary of our results and an outlook on future developments and possible improvements. 

\section{MC-CGC: event generator based on CGC}\label{sec:mc-cgc}

The MC-CGC framework~\cite{Deng:2014vda,MCCGCv20} provides a Monte Carlo implementation of %forward \YN{kt-fact. can apply at mid-rapidity} 
high-energy particle production based on factorized cross sections derived within the CGC effective theory, together with non-perturbative processes. In this section, we describe its realization based on the PYTHIA event generator framework.

\subsection{Structure of the event generator}

\begin{figure}[tbh]
    \centering
        \includegraphics[width=0.42\textwidth]{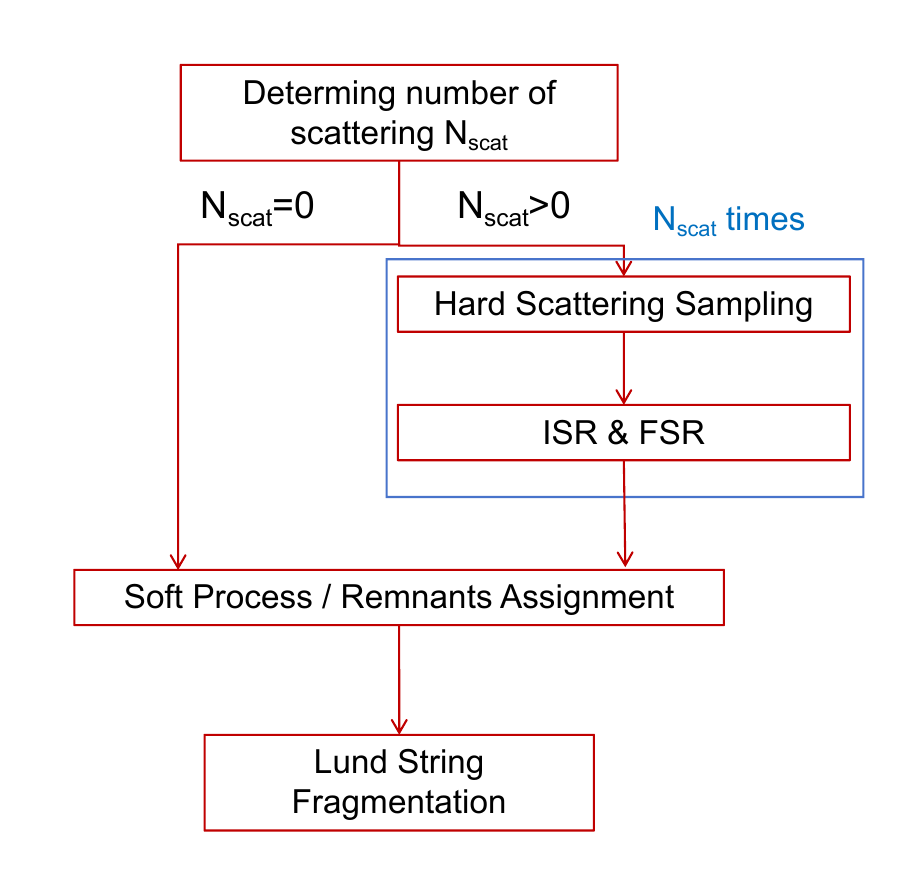}
    \caption{Workflow of the MC-CGC model~\cite{Deng:2014vda} for each event. 
    %\YN{assign remnants kinematics is included Nscat times?}
    \label{fig:workflow}}
\end{figure}

The overall structure of the MC-CGC model for each event is presented in Fig.~\ref{fig:workflow}. At the beginning of the event generation, we determine the number of hard scattering processes $N_{\rm scat}$ sampled from the negative binomial distribution (NBD), with the mean number of scatterings $\langle n\rangle=\sigma T_{pp}(b)$ determined by the impact parameter $b$ dependent thickness function $T_{pp}(b)$ and the total cross section $\sigma$ calculated from the given CGC type factorization formula discussed below. If there is no hard scattering process, we treat the event within the quark exchange model: A valence quark is taken from the projectile (or target), and a color string is formed between this quark and the remaining diquark in the target (or projectile). This process results in two color-singlet strings. In contrast, when the system undergoes at least one scattering process ($N_{\rm scat}\ge 1$), we treat each scattering process independently within the CGC framework outlined below. The scattered partons and generated partons then go through the initial- and final-state radiation processes (ISR, FSR), respectively. After the radiation process is complete, we determine the beam remnant kinematics and connect them to the generated partons to form the color singlet strings. Finally, we group the generated color singlet strings in PYTHIA8.315~\cite{Bierlich:2022pfr} to perform the Lund string fragmentation. 

\subsection{Single particle production formula}

By default, the framework employs the DHJ formula~\cite{Dumitru:2005gt, Deng:2014vda, Duraes:2014jxa}, which describes the scattering of a dilute projectile parton off a dense target within the CGC effective theory. The corresponding parton-level cross-section is given by~\cite{Dumitru:2005gt}
\begin{align}\label{eq:DHJ}
    \frac{\d\sigma_{\rm DHJ}}{\d y\d^2 p_T}=\frac{K}{(2\pi)^2}\frac{\sigma_0}{2}\sum_{i = q, \bar q, g}x_1 f_{i/p}(x_1,Q^2)N_i(x_2, p_T)\, , 
\end{align}
where the rapidity $y$ and the transverse momentum $p_T$ are related to the Bjorken-$x$ by $x_{p,t}=(p_T/\sqrt{s})e^{\pm y}$, with $x_1=\max\{x_p,x_t\}$ and $x_2=\min\{x_p,x_t\}$. The parameter $K$ is introduced as an overall factor, and $\sigma_0/2=16.5\rm\ mb$~\footnote{We note in Ref.~\cite{Lappi:2013zma}, the average transverse area is fitted to be $\sigma_0/2=16.36~\rm mb$ for the MV$^e$ model, while in this work we keep it the same as the fitting in Ref.~\cite{Albacete:2012xq}. The discrepancy will be compensated by the overall normalization $K$.} represents the average transverse area of the proton, constrained by the DIS fits in HERA~\cite{Lappi:2013zma, Albacete:2010sy, Albacete:2009fh}.
Here, $f_{i/p}(x_1, Q^2)$ denotes the collinear parton distribution functions (PDFs) for quarks ($i=q, \bar q$) and gluons ($i=g$) in the projectile proton, 
while $N_i= N_{F, A}$ represents the corresponding dipole amplitudes of quarks (fundamental representation) and gluons (adjoint representation) traversing the dense target, respectively. 
By default, we set the factorization scale $Q$ for the PDFs to $p_T$: $Q=p_T$.

When both the projectile and the target are dense, we use the $k_T$-factorization formalism as a phenomenological model incorporating saturation effects.
%
%\footnote{%\blue{Modifications needed from Fujii-san} Although both projectile and target are described by small-$x$ evolution including saturation effects, the kinematic region considered here satisfies $p_T> Q_s$. In this limit, nonlinear multiple-scattering contributions are suppressed, and the cross section can be approximated by such a $k_T$-factorized form, with transverse momentum dependence involved effectively by the unintegrated gluon distributions. This should be understood as a linearized limit of the scattering process of two CGC fields, in which higher-point correlators are neglected.
%The dense regime is parametrically characterized by $gA \sim O(1)$, where $g$ is the gauge coupling and $A$ the gauge field. Thus, the calculation of gluon production in dense–dense collisions generally requires solving the classical Yang–Mills equations with all-order rescattering effects from both the projectile and target fields. 
%However, for gluon production at $p_T \gg Q_s$, higher-order rescatterings are suppressed by powers of $Q_s^2/p_T^2$.
%In this limit, the cross section can be approximated by the $k_T$-factorized form, while saturation effects remain encoded in the incoming unintegrated gluon distributions $\phi$.
%Nevertheless, in the present work we adopt the $k_T$-factorization form phenomenologically also in the lower-$p_T$ region, expecting it to provide a useful effective description beyond its strict asymptotic validity.}
%
In this approach, the cross-section is given by~\cite{Levin:2010dw}
\begin{align}\label{eq:kt1}
    \frac{\d\sigma_{\rm k_T}}{\d y\d^2p_T}=&\frac{K}{2C_F}\frac{1}{p_T^2}\left(\frac{\sigma_0}{2}\right)^2\int\d^2\kt\ \alpha_s(Q_M^2)\notag\\&\times\phi\left(x_1,\frac{\pt+\kt}{2}\right)\phi\left(x_2,\frac{\pt-\kt}{2}\right)\, , 
\end{align}
where $C_F=\frac{N_c^2-1}{2N_c}$ and $Q_M^2=\max\{(\pt+\kt)^2/4,\ (\pt-\kt)^2/4\}$, and the unintegrated gluon distributions $\phi$ are related to the dipole amplitude via
\begin{align}\label{eq:kt2}
    \phi(x,\kt)=\frac{C_F}{(2\pi)^3}\frac{1}{\alpha_s(Q_s^2(x))}k_T^2 N_A(x,k_T).
\end{align}
The running coupling constant in the unintegrated gluon distribution $\phi(x,\kt)$ is taken as $\alpha_s(q^2) = \text{min}(\alpha_{\text{fr}}, 4 \pi/\beta_0  \ln(q^2/\Lambda^2) )$ with $\alpha_\text{fr} = 0.5$ and $\Lambda=0.24$ GeV. The choice of $Q^2$ scale in Eqs.~\eqref{eq:kt1} and \eqref{eq:kt2} follows~Ref.~\cite{Levin:2010dw}.

\begin{figure}[tbh]
    \centering
        \includegraphics[width=0.42\textwidth]{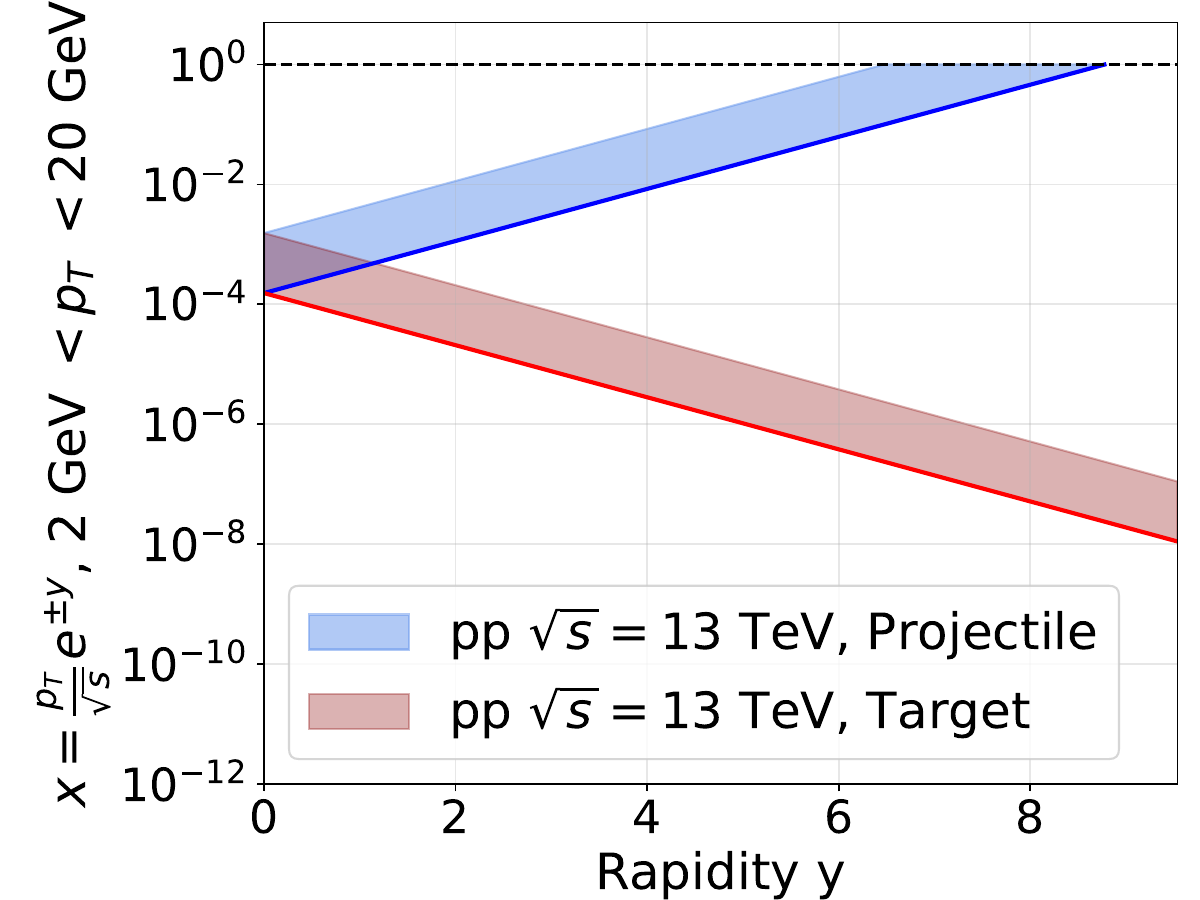}
    \caption{Coverage of the Bjorken-$x$ range for $2$ GeV $< p_T < 20$ GeV in pp collisions at $\sqrt{s} =13$ TeV. The horizontal dashed line sets 
    the upper bound $x = 1$. The blue band and red band are the coverage $x$-range for projectile and target, respectively, with their lower limits defined by the corresponding solid lines.}\label{fig:xrange}
\end{figure}

It is important to assess the predictive power of the dilute--dense description by comparing it with a dense--dense description. The motivation for this comparison is illustrated in Fig.~\ref{fig:xrange}, which shows the Bjorken-$x$ ranges accessed in the pp collisions at 13 TeV. The system can be divided into distinct rapidity regions, each characterized by different dynamics.

%%%%%%%%

At central and mid-rapidity ($y<2$) at the LHC energies, both projectile and target partons reside in the small-$x$ regime, corresponding to a dense--dense scattering environment. 
In this region, the predictions of the DHJ formula — which relies on collinear factorization and leading-order PDFs, may become less reliable. 
In contrast, the $k_T$-factorization framework consistently incorporates these saturation effects on both sides and is therefore expected to provide a more reliable description. 

In the intermediate forward region ($2<y<6$), the projectile parton has values of $x_p$ larger than $10^{-2}$, while the target parton reaches at smaller values of $x_t$. Since the rcBK evolution in our model is initialized at $x_0=10^{-2}$ and then performed toward smaller values of $x$, the $k_T$-factorization result has to rely on the extrapolation of the dipole amplitude at $x_p>x_0$, which reduces its predictive accuracy. On the other hand, the use of collinear PDFs for the projectile with larger values of $x_p$ becomes more justified, while the target is probed at smaller values of $x_t$, reinforcing the dilute--dense asymmetry. As a result, the DHJ formalism becomes more applicable in this intermediate forward region. 

At even larger rapidities ($y>6$), the projectile is sufficiently dilute that the DHJ formula should reliably capture the dominant saturation physics from the target. However, as $y$ approaches the beam rapidity, non-perturbative beam-remnant effects may become significant and could mimic or obscure genuine CGC signals.

%\red{In the present calculation, we use the NNPDF3.1 leading-order (LO) parton distribution functions~\cite{NNPDF:2017mvq} as the default input for the projectile proton, and the dipole amplitudes $N_{i}$ constructed from the solutions of the running-coupling Balitsky--Kovchegov (rcBK) evolution equation. This choice is motivated by the fact that the particle-production cross section is evaluated using the LO DHJ formalism, so the use of LO PDFs provides a consistent baseline within the present implementation. The calculation is therefore not intended as a full next-to-leading-order (NLO) analysis. } 

In the present calculation, we use the NNPDF3.1 leading-order (LO) parton distribution functions~\cite{NNPDF:2017mvq} as the default input for the projectile proton, and the dipole amplitudes $N_{i}$ constructed from the solutions of the running-coupling Balitsky--Kovchegov (rcBK) evolution equation. This choice is motivated by the fact that the particle-production cross section is evaluated using the LO DHJ formalism, so the use of LO PDFs provides a consistent choice within the present implementation. The initial condition for the rcBK evolution is set at $x_0=10^{-2}$, and we study three different types of initial condition (McLerran--Venugopalan (MV) model~\cite{McLerran:1993ni, McLerran:1993ka}, 
and its variants, MV$^\gamma$~\cite{Albacete:2012xq} and MV$^e$~\cite{Lappi:2013zma}). The initial dipole amplitudes in the coordinate space $r\sim 1/k_T$ are represented by 
\begin{align}
    N_F(x_0,r)&=1-\exp\left[-\frac{(r^2Q_{s0}^2)^\gamma}{4}\ln\left(\frac{1}{\Lambda r}
    + e_c\cdot e\right)\right]\, ,
\end{align}
where $\Lambda=\Lambda_{\rm QCD}=0.241$ GeV and the parameters $\gamma$, $Q_{s0}$, and $e_c$ characterize each initial condition. The running coupling constant in the coordinate space is taken as 
$\alpha_s(r) = 4\pi / \left[\, \beta_0 \ln \left( \frac{4C^2}{r^2 \Lambda^2} + \mu \right) \,\right]$, where $\mu$ is fixed by the condition $\alpha_s(\infty) = \alpha_{\rm fr}$. 

\begin{figure*}[hbt]
    \centering
        \includegraphics[width=0.42\textwidth]{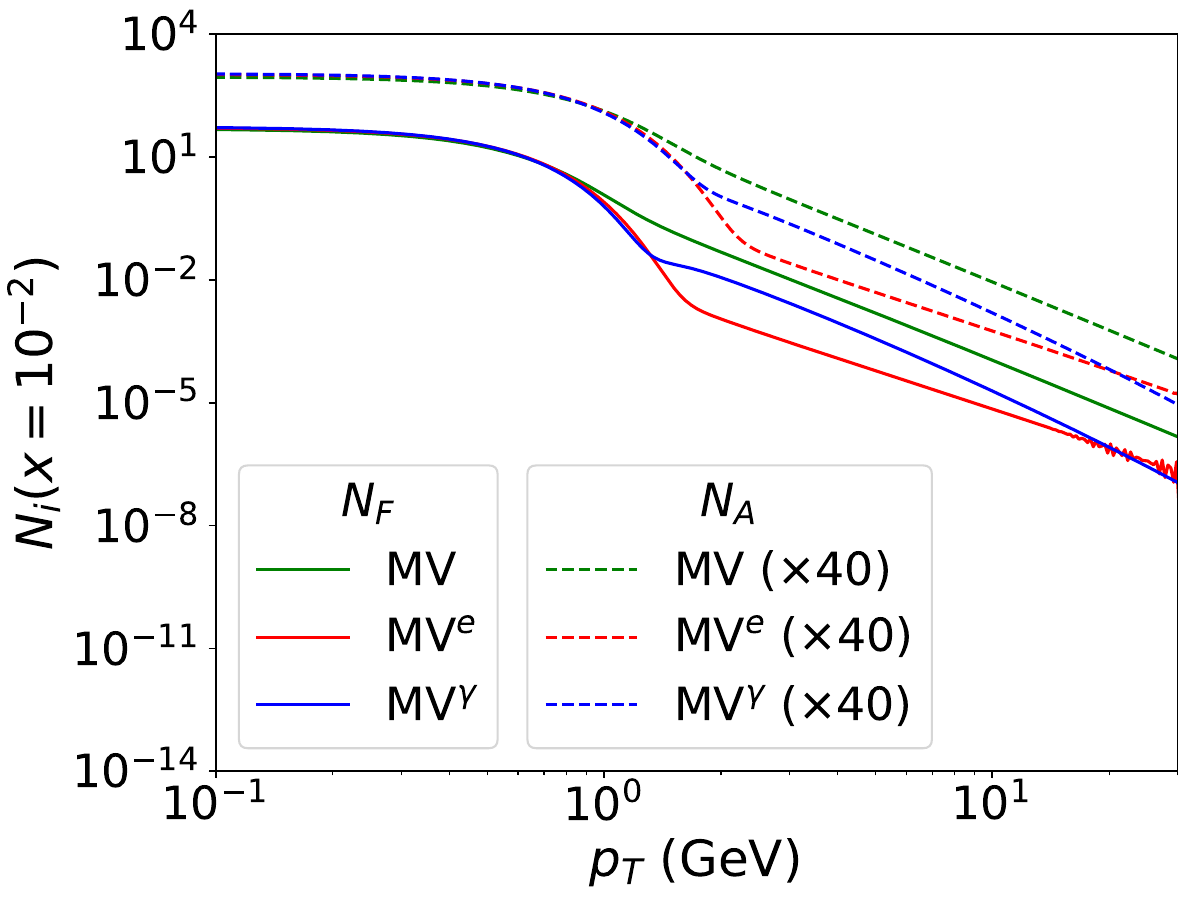}
    \hspace{0.5cm}
        \includegraphics[width=0.42\textwidth]{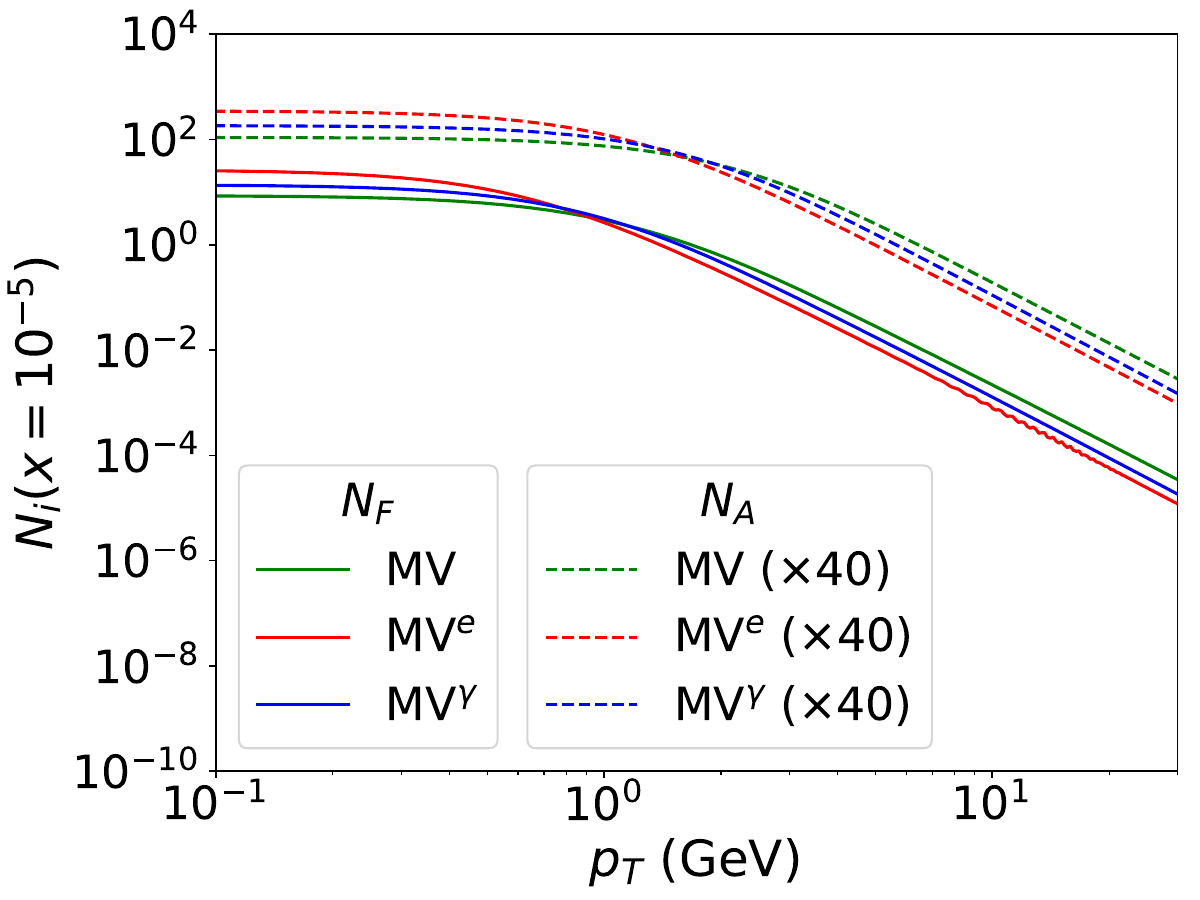}
    \caption{Dipole amplitude from three initial conditions used in this paper: MV model (green), MV$^e$ model (red) and MV$^\gamma$ model (blue). The left panel shows the initial conditions for the dipole amplitude defined at $x_0=0.01$, and the right panel shows the evolved dipole amplitude at $x=10^{-5}$. Solid lines represent the dipole amplitude in the fundamental representation, while dashed lines are for the adjoint representations. \label{fig:dipole}}
\end{figure*}

The parameters for these initial conditions are summarized in Table \ref{tab:input_parameters}. 
We mainly use the g1.101 parameterization from Ref.~\cite{Albacete:2012xq}, generated with the MV$^\gamma$ initial condition. The shape of dipole amplitudes from the three parameterizations is displayed in Fig.~\ref{fig:dipole}. We observe that at the beginning of the evolution ($x=x_0$), the three initial conditions yield very different structures, with a knee structure appearing in the MV$^\gamma$ and MV$^e$ type initial conditions. As the dipole amplitude evolves to the forward rapidity, the knee structure disappears, and the evolution of the three initial conditions shares a similar shape, reflecting the attractor behavior of the rcBK evolution.

\begin{table}[h]
    \centering
    \begin{tabular}{cccccc}
    \hline
    \hline
        rcBK ICs    & $\gamma$  & $Q_{s0}^2$ (GeV$^2$)  & $C^2$ & $e_c$ & $\alpha_{\rm fr}$\\
   \hline
        MV          & $1$       & $0.200$               & $1$           & 1  & 0.5 \\
        MV$^\gamma$ & $1.101$   & $0.157$               & $1$           & 1  & 0.8 \\
        MV$^e$      & $1$       & $0.060$               & $7.2$         & $18.9$ & 0.7 \\
    \hline
    \hline
    \end{tabular}
    \caption{Parameters for rcBK initial conditions taken from Refs.~\cite{Albacete:2012xq, McLerran:1993ni, McLerran:1993ka, Lappi:2013zma}. Note
    $\alpha_s(\infty) = \alpha_{\rm fr}$ is controlled by $\mu$ for MV and MV$^\gamma$, while
    $\mu=0$ and $\alpha_s(r)$ is frozen at 0.7 in the IR region for MV$^e$.}
    \label{tab:input_parameters}
\end{table}

In this work, we also investigate theoretical uncertainties stemming from various inputs. The dependence on collinear PDFs is tested with MV$^\gamma$ initial condition using the CT18LO~\cite{Yan:2022pzl} ($K=1.8$), HERA20LO~\cite{H1:2015ubc} ($K=1.5$), and NNPDF3.1 NLO~\cite{NNPDF:2017mvq} ($K=1.5$) sets in addition to the default NNPDF3.1 LO~\cite{NNPDF:2017mvq} ($K=1.4$). We find that the differences among the leading-order PDF sets can be largely absorbed by the $K$-factor, leading to minimal sensitivity in the final spectra. The test of the NNPDF3.1 NLO PDF set also gives consistent results on forward hadron production, while some differences appear close to mid-rapidity at 13 TeV. Furthermore, we explore the sensitivity to the initial conditions for the rcBK evolution by considering the MV and MV$^e$ models alongside the default MV$^\gamma$ initial condition.

\subsection{Multi-parton hard scatterings}

Multiple hard scatterings can occur within a single proton--proton event at the LHC energies. 
We allow the number of hard scatterings to fluctuate on an event-by-event basis such that the resulting charged-particle multiplicity distribution follows a negative binomial distribution~\cite{Dumitru:2012yr}, as suggested in  Refs.~\cite{UA1:1982yyh, UA5:1985hzd, UA5:1985kkp, UA5:1985fid}. These fluctuations are implemented through the following procedure.

First we fix the mean number of hard scatterings to $\langle n \rangle= \sigma T_{pp}(b)$,
where $\sigma$ denotes either the DHJ \eqref{eq:DHJ} or the $k_T$-factorization \eqref{eq:kt1} cross-section described above,
and the impact parameter dependent thickness function $T_{pp}(b)$  is assumed to have a Gaussian form with variance controlled by the free parameter $B$,
\begin{align}
    T_{pp}(b)=\frac{1}{4\pi B}\exp\left(-\frac{b^2}{4B}\right).
\end{align}
Next, to generate the negative binomial distribution of the number of hard scatterings $n$, 
\begin{align}\label{eq:nbd}
P(n) = \frac{\Gamma(k+n)}{\Gamma(k)\Gamma(n+1)}
\frac{ {\langle n \rangle}^n k^k }{( \langle n \rangle + k)^{n+k}}\, ,
\end{align}
We employ a two-step procedure as discussed in Refs.~\cite{Bozek:2013uha, Moreland:2014oya}. An auxiliary variable $\lambda$ is first sampled from a Gamma distribution $\Gamma(k, \theta)$ with shape and scale parameters $k$ and $\theta$, chosen such that $k \theta = \langle n \rangle$. Subsequently, the number of scatterings $n$ in each event is drawn from a Poisson distribution with mean $\lambda$ determined above. The resulting distribution of $n$ is a negative binomial distribution with mean $\langle n \rangle$ and the shape parameter $k$. %The parameters $B=0.325\rm\ fm^2$ and $k=0.9$ are fixed by fitting the LHCb data.
%\YN{how do you find this method? Is this the standard method? If you took from somewhere, add the reference. If not, it is hard to understand this method. It would be good to write more and put it in the appendix.}

The equivalence of the two-step procedure and NBD is shown below. Consider the Poisson distribution and the Gamma distribution defined by
\begin{align}
    &\mathrm{Poisson}(n|\lambda)=\frac{e^{-\lambda}\lambda^n}{n!},\\
    &\mathrm{Gamma}(\lambda|\theta,k)=\frac{1}{\Gamma(k)\theta^k}\lambda^{k-1}e^{-\lambda/\theta}.
\end{align}
Their marginal distribution is
\begin{align}
    P(n)  =&\int_0^\infty \mathrm{Poisson}(n|\lambda)\, \mathrm{Gamma}(\lambda|\theta,k) d\lambda \notag\\
    &=\frac{1}{n!\Gamma(k)\theta^k}\int_0^{\infty}\lambda^{n+k-1}\exp\left[-\lambda\left(1+\frac1\theta\right)\right]\d\lambda\notag\\
    &=\frac{\Gamma(n+k)}{\Gamma(n+1)\Gamma(k)}\left(\frac{\theta}{1+\theta}\right)^n\left(\frac{1}{1+\theta}\right)^k.
\end{align}
Taking $\theta=\langle n\rangle /k$ reproduces Eq.~\eqref{eq:nbd}.

\begin{comment}
\com{HF: Which work first did this two-step sampling?}

\blue{For the two-step sapling, Eq.8 in ~\cite{Bozek:2013uha} and the discussion below eq.12 in ~\cite{Moreland:2014oya} have explained the motivation, and this sapling procedure is implemented in ~\cite{Moreland:2014oya}, which has been well used as the initial condition for modeling heavy-ion collisions. Further, it's possible to directly prove it. Consider the Poisson distribution and the Gamma distribution
\begin{align}
    \mathrm{Poisson}(n|\lambda)=\frac{e^{-\lambda}\lambda^n}{n!},\ \ \mathrm{Gamma}(\lambda|\theta,k)=\frac{1}{\Gamma(k)\theta^k}\lambda^{k-1}e^{-\lambda/\theta}
\end{align}
Then what I used is the marginal distribution:
\begin{align}
    P(n)  =&\int_0^\infty \mathrm{Poisson}(n|\lambda)\, \mathrm{Gamma}(\lambda|\theta,k) d\lambda \notag\\
    &=\frac{1}{n!\Gamma(k)\theta^k}\int_0^{\infty}\lambda^{n+k-1}\exp\left[-\lambda\left(1+\frac1\theta\right)\right]\d\lambda\notag\\
    &=\frac{\Gamma(n+k)}{\Gamma(n+1)\Gamma(k)}\left(\frac{\theta}{1+\theta}\right)^n\left(\frac{1}{1+\theta}\right)^k
\end{align}
Define $p=1/(\theta+1)$, we have
\begin{align}
    P(n)=\frac{\Gamma(n+k)}{\Gamma(n+1)\Gamma(k)}p^k(1-p)^n
\end{align}
which is the negative binomial distribution. In Nara-san's paper~\cite{Dumitru:2012yr}, they directly use the distribution, and here by defining $p=k/(\bar{n}+k)$ with $\bar{n}=\sigma T_{pp}(b)=\theta k$ in our paper, the two expressions describe the same distribution.
}
\end{comment}

In the current model, with the scattering process sampled from the DHJ formula, events with multiple hard parton scatterings are restricted to involve at most one quark-initiated process, with the remaining scatterings assumed to be gluon-initiated. This treatment is used to simplify the treatment of color reconnection (described below) and will be improved in the next update, aiming to extend to pA collisions. In the $ k_T$-factorization formula, all hard scatterings are initiated by gluons, and no further improvement from quark contributions is needed.

\subsection{Initial and final state radiations}
Once the hard scattering has been determined, initial- and final-state parton radiations are subsequently simulated. Here, we restrict the model to account for initial-state radiation (ISR) from the dilute projectile parton using a backward-evolution algorithm~\cite{Sjostrand:1985xi, Marchesini:1987cf, Sjostrand:2006za, Sjostrand:2007gs}. Radiation associated with the dense target is not generated as an exclusive shower; instead, it is effectively resummed into the rcBK-evolved gluon distribution. This treatment captures the dominant small-$x$ dynamics of the target, while explicit radiation effects are expected to be subleading for forward particle production. By contrast, we always perform the final-state radiation (FSR) from the produced hard parton by using the forward branching algorithm~\cite{Sjostrand:2006za, Bierlich:2022pfr}. %For both ISR and FSR, a minimal $p_T$ cut $p_{T,\mathrm{rad},\min}=Q_{\rm cut,\ min}/2=1.5\rm\ GeV$ is applied. %As we have checked, a larger regulator, compared with the PYTHIA setting, can produce consistent results when replaced by the auxiliary scale $\Lambda$ discussed in ~\cite{Shi:2021hwx}. 

%%%%%%%%%%%%%%%%%%%%%%%%%%%%%%%%%%%%%%%%%%%%%%
\subsection{Remnants}

To construct color-singlet systems from partons in the Lund string fragmentation model, beam remnants are included as well. The remnant momentum is fixed by overall momentum conservation, serving as a recoil partner for the remaining system. When a quark $q$ comes from the projectile or target, we assume it is always connected to the diquark pair. In contrast, when a gluon $g$ is extracted from the projectile, the remnant system breaks into a quark $q_1$ and a diquark, $(q_2 q_3)$. These remnants are then connected to the generated partons from scattering and radiation process to form the string with the shape $q_1 - g-\cdots -g- (q_2  q_3)$~\cite{Wang:1991hta}. The remnant splitting procedure can occur on both the projectile and target sides, and the generated gluons are attached to the remnants. The energy fraction $\chi$ carried by one of the two remnants is sampled from the distribution
\begin{align}
    P(\chi) \propto\frac{(1-\chi)^k}{(\chi^2+c_{\rm min}^2)^{b_\chi/2}}, 
\end{align}
where PYTHIA default values $k=3$ and $c_{\rm min}= 2\langle m_q\rangle / E_{\rm cm}=(0.6~{\rm  GeV})/E_{\rm cm}$ are taken, and $b_\chi=1$ matches the setting of PYTHIA8~\cite{Bierlich:2022pfr}.  By varying the parameters $k\in[1,3]$ and $b_\chi\in[0.5,1.5]$, we find little effect on the final observables in the considered pseudo-rapidity acceptance $|\eta|<6$. With $\chi$ determined, the light-cone momenta $p^{\pm}$ of the breaking remnants can be reconstructed.  The transverse recoil momentum is split equally between the two remnants, with additional intrinsic Gaussian-type momentum fluctuations with the width of $2\rm\ GeV$, which is the same as the default PYTHIA setting.

\subsection{Fragmentation}
The generated partons and remnants are hadronized via the Lund string fragmentation model implemented in PYTHIA 8.315~\cite{Bierlich:2022pfr}. The identified hadron species considered in this work, including pions, $\eta$, and $\omega$ mesons, are generated through this hadronization and subsequent hadronic decay procedure. The fragmentation function for splitting a string into hadrons of momentum fraction $z$ follows the Lund form:
\begin{align}
    f(z) = (1/z)(1-z)^a\exp(-b m_T^2/z),
\end{align}
where $m_T^2=m_q^2 + p_T^2$, with quark transverse momentum $p_T$ sampled from Gaussian distribution with width $\sigma_w$. The default Lund string parameters $a=2.0$ and $b=0.2$ are used in this calculation.
%At each string break, the produced quark--antiquark pair receives a transverse momentum kick
%to count for the non-perturbative effect, 
%which is sampled from a Gaussian distribution.
%with width $\sigma=0.45\rm\ GeV$. 
%We have verified that the modifications to the $p_T$ spectrum from the Lund $b$ parameter and the $p_T$ kick predominantly affect the spectrum in the soft region ($p_T\le 3\rm\ GeV$), while their impact in the hard region can be effectively compensated by adjusting the overall $K$-factor.

\subsection{Parameter setting}

In this work, most of the parameters are fixed either by using fits to the DIS data from HERA or by adopting the default values suggested by PYTHIA~\cite{Sjostrand:2006za, Bierlich:2022pfr}. A subset of parameters, however, is tuned to improve agreement with existing experimental data. When we determine the number of hard scatterings, the Gaussian width parameter $B=0.325\rm\ fm^2$ in the thickness function and the shape parameter $k=0.9$ are chosen to reproduce the multiplicity distribution measured by LHCb~\cite{LHCb:2014wmv}. The fluctuations reflected by $k$ may originate from subnucleonic structure or intrinsic fluctuations of the saturation scale $Q_s^2$, as discussed in~\cite{McLerran:2015qxa}. For scattering and radiation processes, the minimum transverse momentum cut is set to $p_T = 1.5~\rm GeV$, corresponding to a larger infrared regulator. We have verified that this choice produces consistent results compared with using the auxiliary scale $\Lambda$ as discussed in~\cite{Shi:2021hwx}. In the fragmentation stage, the Gaussian width of the transverse momentum kick is increased from $\sigma_w=0.35~\rm GeV$ to $\sigma_w = 0.45~\rm GeV$ to improve the description at mid-rapidity. In the forward region, changes in $\sigma_w$ are compensated by adjusting the overall normalization or by effectively mimicking NLO contributions, leaving room for further refinement in future studies.

%%%%%%%%%%%%%%%%%%%%%%%%%%%%%%%%%%%%%%%%%%%%%%%%%%%%%%%%%%%%%%%%%%%%%%%%%%%%%%%%%%%%%%%%
\section{Results}
\subsection{Evaluation of Existing Measurements}\label{sec:Evaluation-of-Existing-Measurements}

\subsubsection{Particle production at LHCb}

\begin{figure*}[hbt]
    \centering
        \includegraphics[width=0.42\textwidth]{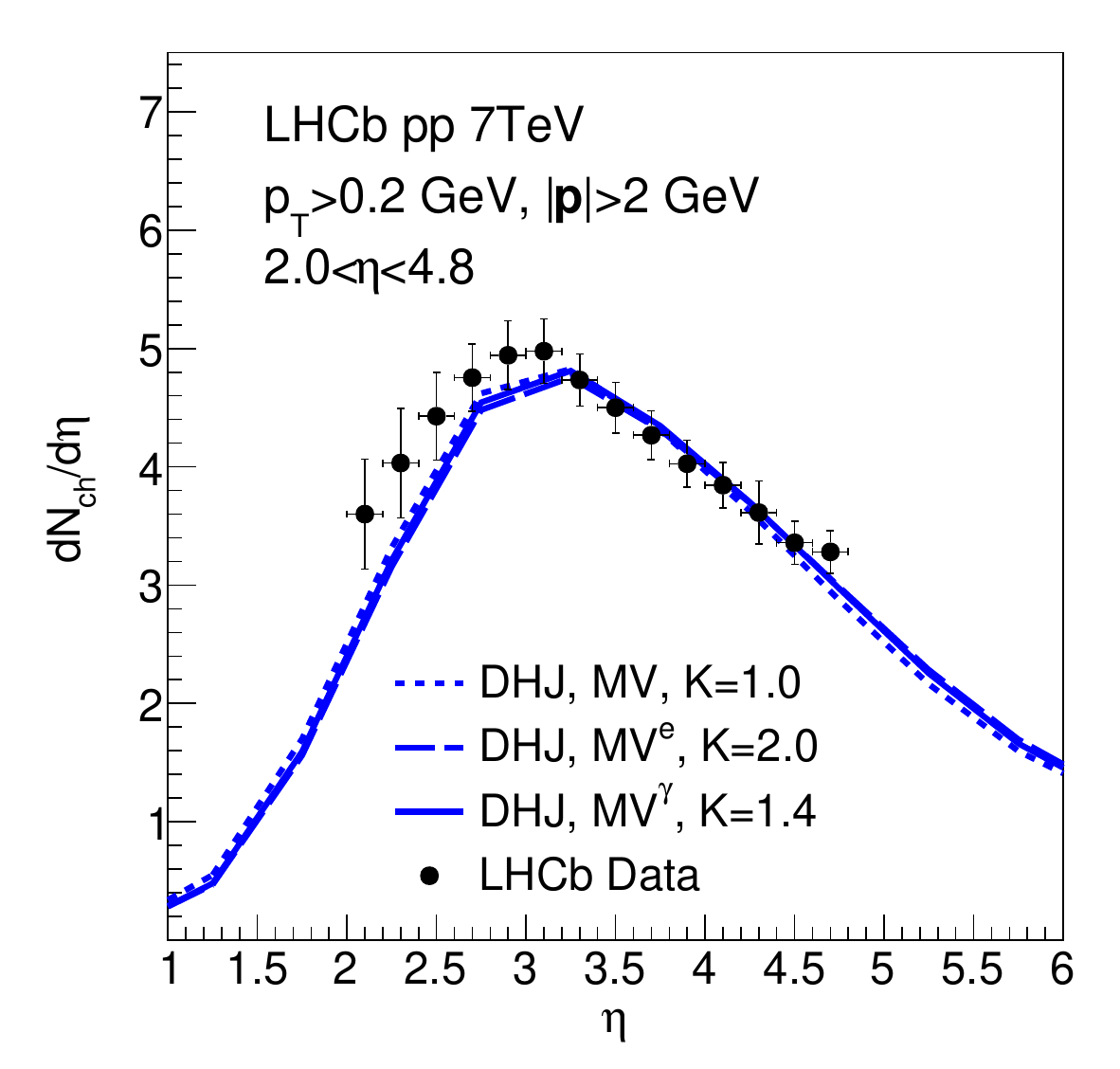}
    \hspace{0.1cm}
        \includegraphics[width=0.42\textwidth]{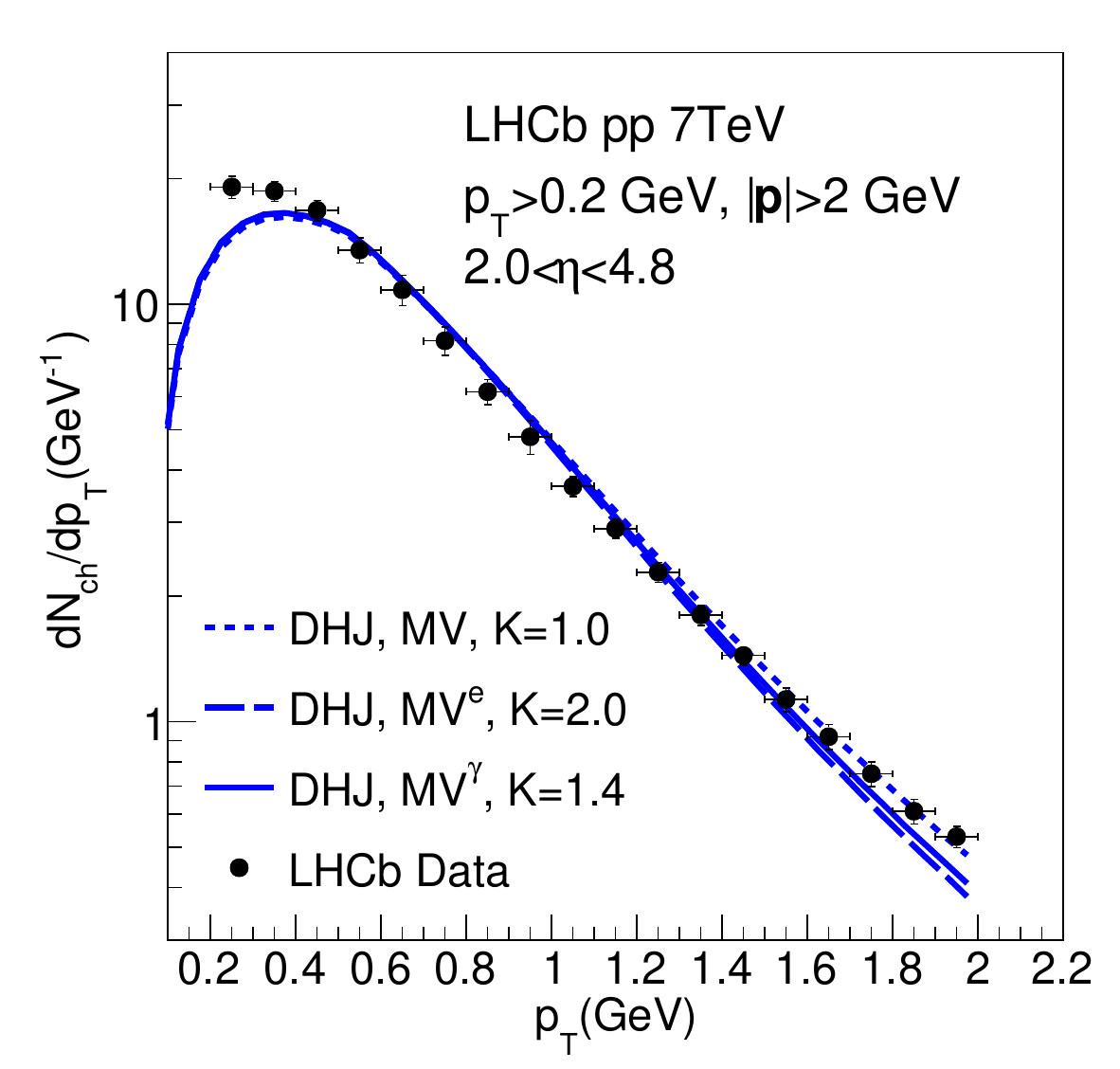}
    \caption{$\d N_{\rm ch}/\d\eta$ (left) and  $\d N_{\rm ch}/\d p_{T}$ (right) in pp collisions at $\sqrt{s}=7\rm\ TeV$. Experimental data (dotted points)  taken from Ref.~\cite{LHCb:2014wmv}. Model simulations are performed based on the DHJ formula with the three different initial conditions and corresponding normalization factor $K$: MV model (dotted lines), MV$^e$ model (dashed lines), and MV$^\gamma$ model (solid lines).}\label{fig:fig4}
\end{figure*}

As our model includes several non-perturbative processes, we calibrate our model parameters using soft-particle production measured within the LHCb acceptance. Figure~\ref{fig:fig4} presents the pseudo-rapidity ($\eta$) and the transverse momentum ($p_T$) distribution of charged particles, $\d N_{\rm ch}/\d\eta$ and $\d N_{\rm ch}/\d p_{T}$ in pp collisions at $\sqrt{s}=7$ TeV. The DHJ formula with three initial conditions (MV, MV$^e$, and MV$^\gamma$) reproduces the data with different normalization factors $K$. 
The description of $\d N_{\rm ch}/\d\eta$ slightly underestimates in the middle region ($|\eta|<3$), whereas it provides a reasonable description in the forward ($|\eta|>3$) region, covering the FoCal acceptance. For the $p_T$ spectra $\d N_{\rm ch}/\d p_{T}$ as shown in the right panel of Fig.~\ref{fig:fig4}, the model describes the data down to the ultra-soft region $p_T \approx 0.5\ \rm{GeV}$, indicating a reasonable modeling of the soft production, including contributions from radiation, fragmentation, and non-perturbative corrections.

\begin{figure*}[hbt]
    \centering
        \includegraphics[width=0.42\textwidth]{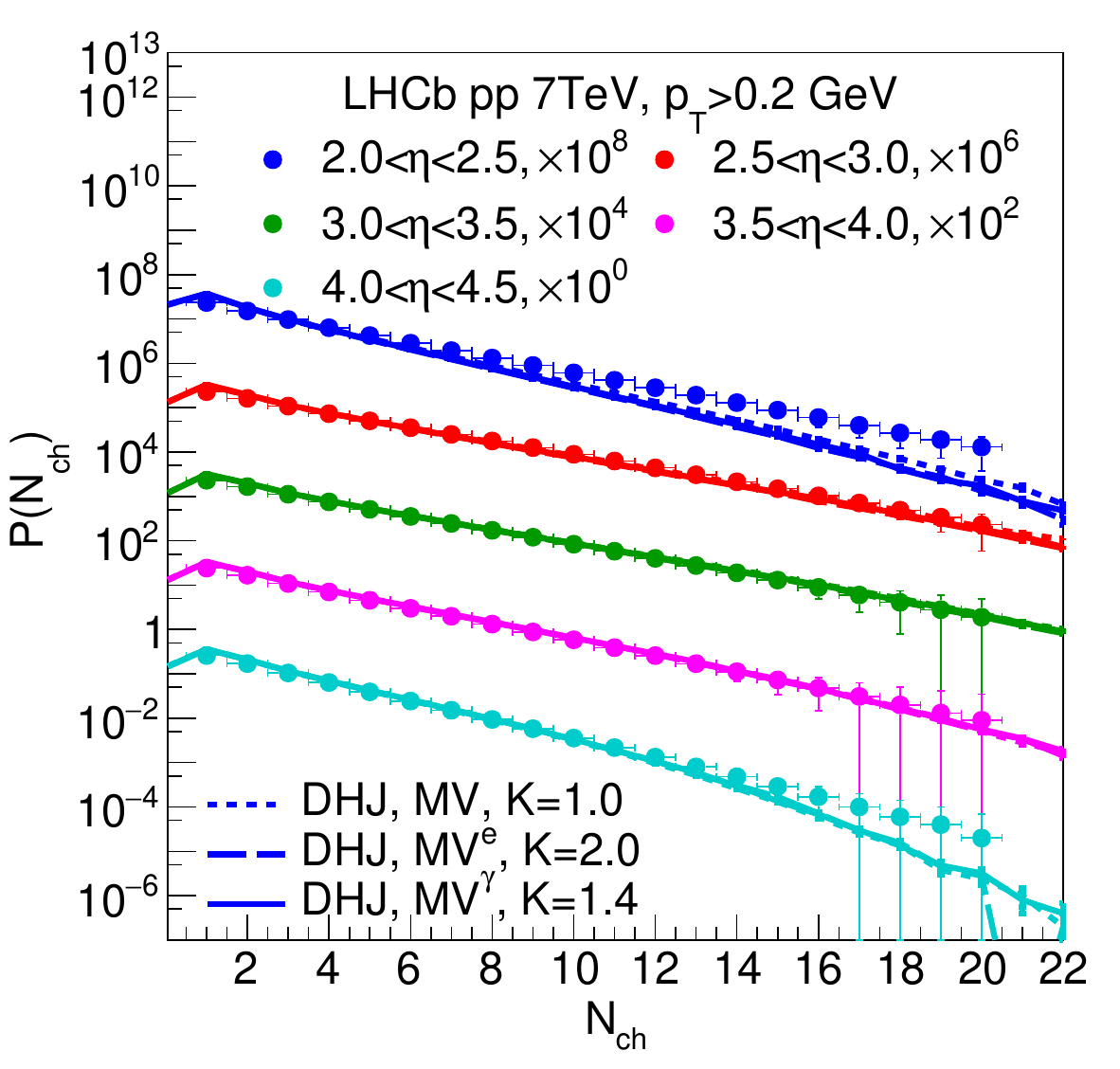}
    \hspace{0.5cm}
        \includegraphics[width=0.42\textwidth]{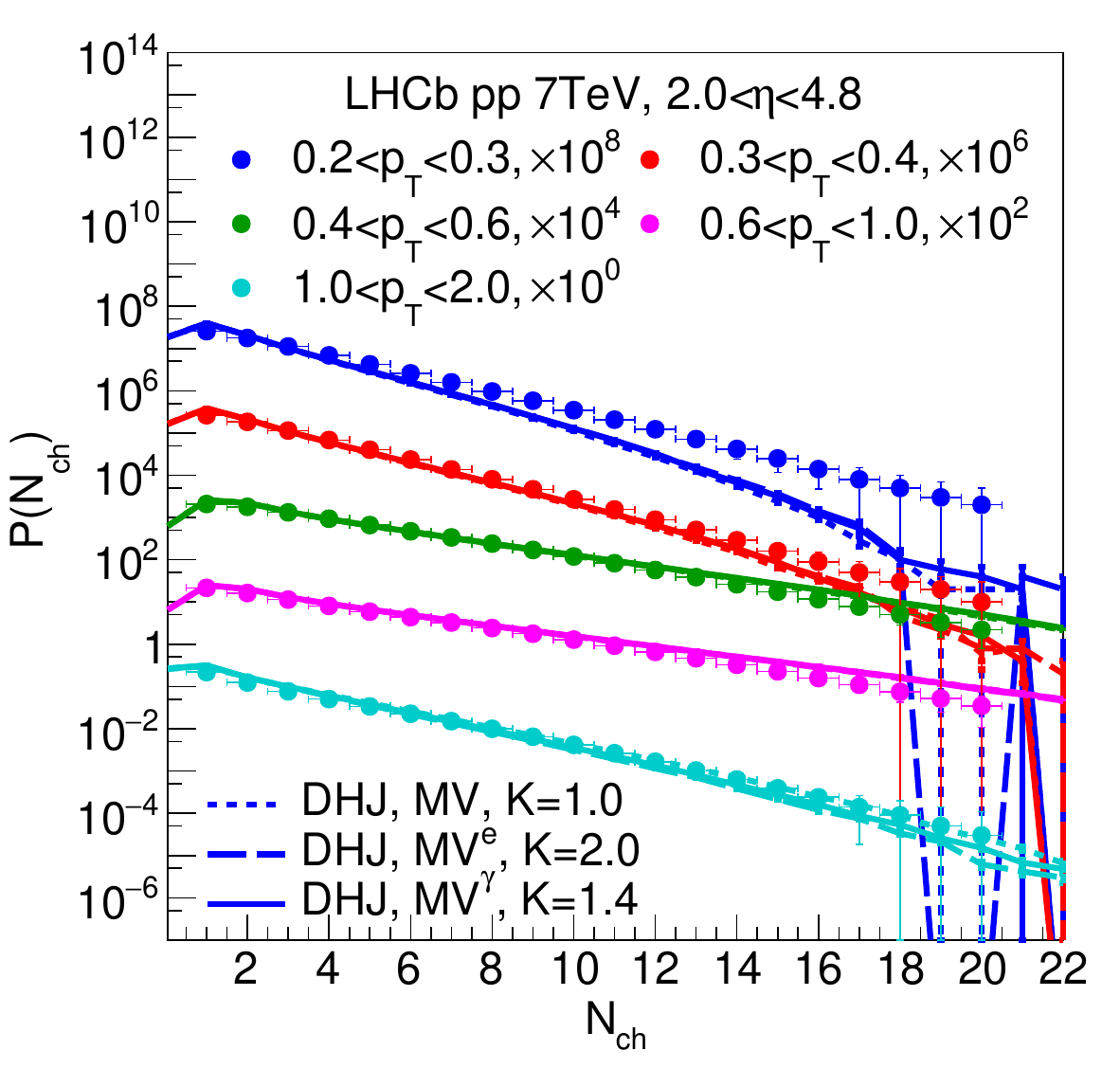}
    \caption{Charged particle multiplicity distribution $P(N_{\rm ch})$ within different $\eta$ (left)  and $p_{T}$ (right) bins in pp collisions at $\sqrt{s}=7\rm\ TeV$. Experimental data taken from Ref.~\cite{LHCb:2014wmv}, with different colors representing different $\eta$ or $p_T$ bins. Model simulations are performed based on the DHJ formula with the three different initial conditions and corresponding normalization factor $K$: MV model (dotted lines), MV$^e$ model (dashed lines), and MV$^\gamma$ model (solid lines).}\label{fig:fig5}
\end{figure*}

The charged-particle multiplicity distributions in different $\eta$ and $p_T$ bins in pp collisions at $\sqrt{s}=7$ TeV are shown in Fig.~\ref{fig:fig5}. The model reproduces the data well in the intermediate $\eta$ region $2.5 < \eta < 4.0$. %\Remove{However, it predicts a narrower distribution at mid- and ultra-forward rapidities, potentially due to an insufficient treatment of beam remnants or the absence of $Q_s^2$ fluctuation.} 
%\red{The multiplicity distribution for different $p_T$ regions shown in the right panel of Fig.~\ref{fig:fig5} is underestimated in the ultra-soft region, but achieves a reasonable description for $p_T > 0.4\ \rm{GeV}$. The deviations at high multiplicities are mainly associated with the underestimated particle yield in the soft region, $0.2~\mathrm{GeV}<p_T<0.4~\mathrm{GeV}$, where the present implementation does not provide a complete description of non-perturbative soft production. Since the present study focuses on semi-hard particle production, this deficiency has a limited impact on the observables discussed below.} 
The multiplicity distribution for different $p_T$ regions shown in the right panel of Fig.~\ref{fig:fig5} is underestimated in the ultra-soft region, but achieves a reasonable description for $p_T > 0.4\ \rm{GeV}$. The deviations at high multiplicities are mainly associated with the underestimated particle yield in the soft region, $0.2~\mathrm{GeV}<p_T<0.4~\mathrm{GeV}$, where the present implementation does not provide a complete description of non-perturbative soft production. Since the present study focuses on semi-hard particle production, this deficiency has a limited impact on the observables discussed below. Calculations with all three rcBK initial conditions agree with each other, suggesting that the model reasonably captures the features of soft production.

\begin{figure*}[hbt]
    \centering
        \includegraphics[width=0.42\textwidth]{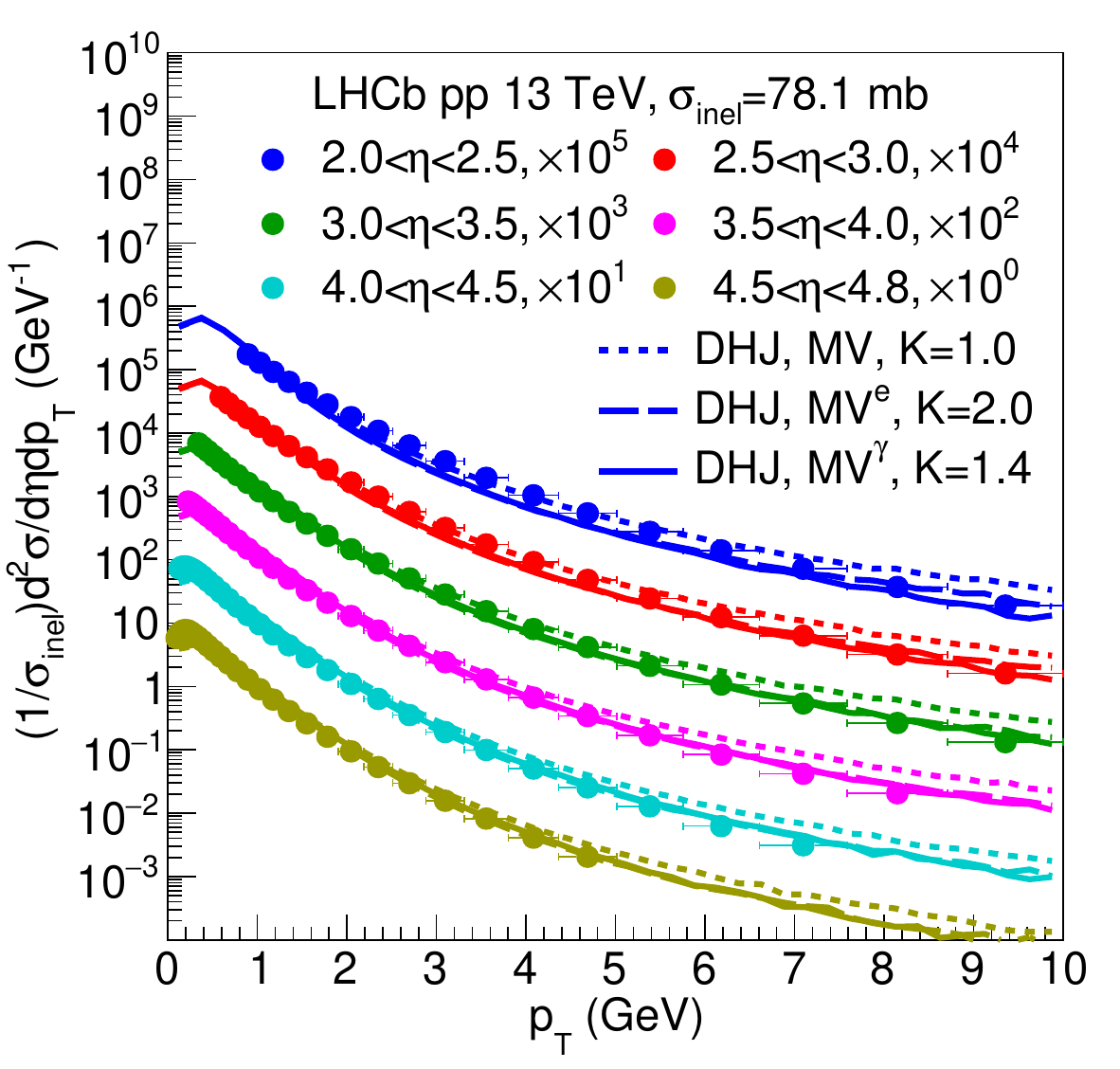}
    \hspace{0.5cm}
        \includegraphics[width=0.42\textwidth]{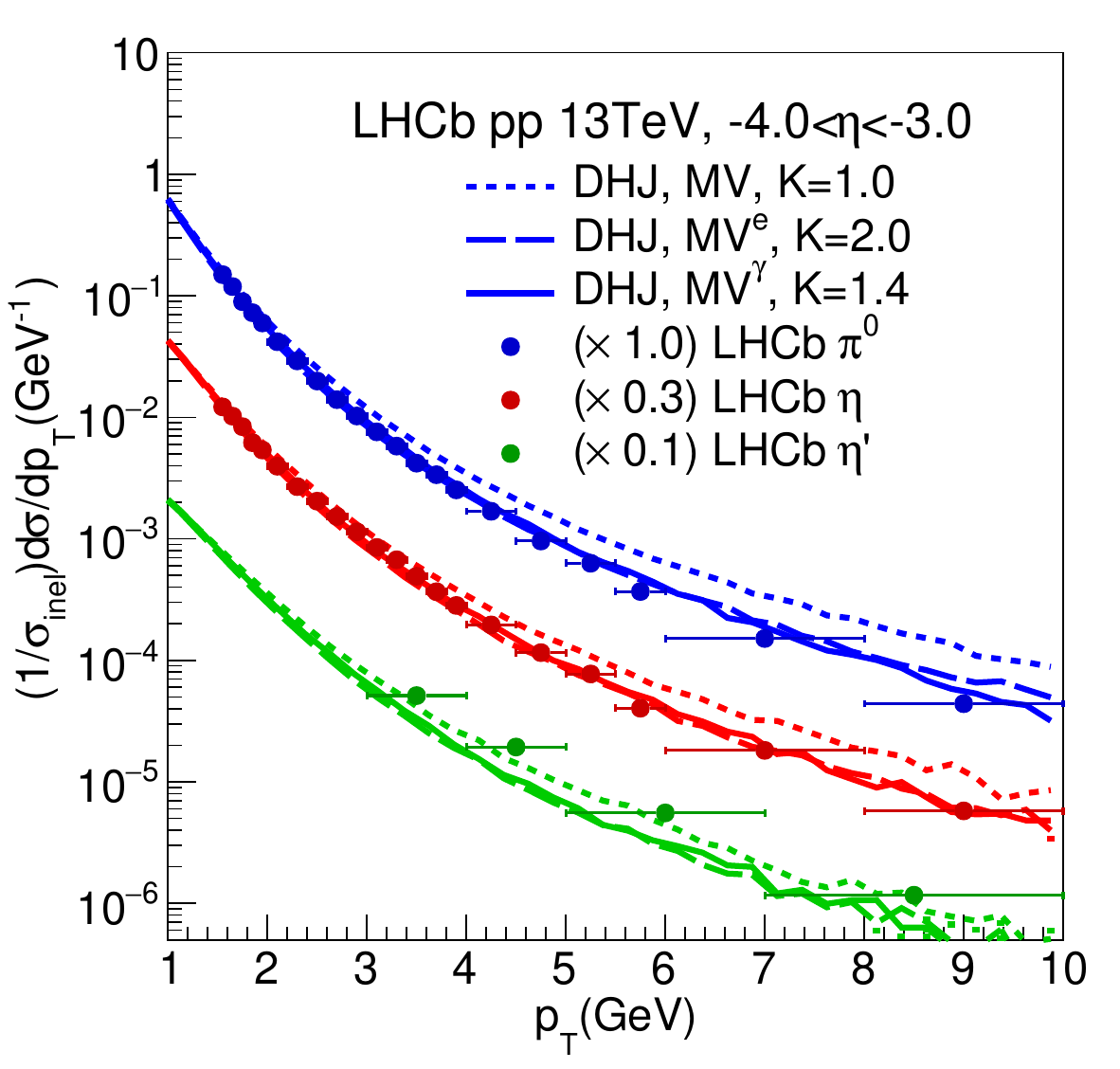}
    \caption{$p_T$ spectra of  charged particles (left) at different $\eta$ bins and $\pi^0$, $\eta$, and $\eta'$ hadrons (right) in pp collisions at $\sqrt{s}=13\rm\ TeV$. Experimental data taken from Refs.~\cite{LHCb:2021abm, LHCb:2023iyw, LHCb:2022tjh}. Model simulations are performed based on the DHJ formula with the three different initial conditions and corresponding normalization factor $K$: MV model (dotted lines), MV$^e$ model (dashed lines), and MV$^\gamma$ model (solid lines).} \label{fig:fig6}
\end{figure*}

With the soft sector reasonably calibrated ($p_T>0.4$ GeV), 
we then investigate the particle production in the hard $p_T$ region.
In the left panel of Fig.~\ref{fig:fig6}, we compare the results of the transverse momentum distributions of charged particles 
with the LHCb data in pp collisions at $\sqrt{s}=13$ TeV.
While all three parameterizations roughly describe the data, the MV model produces a flatter spectrum, disfavored by the experiment. In contrast, the MV$^\gamma$ and MV$^e$ models provide descriptions that are compatible with the measurement. While the differences among the three parameterizations are visible in the hard region near mid-rapidity, they become diminished at forward rapidity.

%\Remove{The right panel of Fig.~\ref{fig:fig6} presents the description for the identified particle production. While the rcBK dipole amplitude is largely flavor-independent, the model shows the different prediction power of $\pi^0$, $\eta$, and $\eta'$ production compared with experiments, which may indicate a coupling between the CGC-type scattering process and the later evolutions.}
We compare the transverse momentum spectra of $\pi^0$, $\eta$, and $\eta'$ with the LHCb data in the right panel of Fig.~\ref{fig:fig6}.  Our prediction with MV$^\gamma$ and MV$^e$ initial conditions is in good agreement with the data, while the calculations from the MV initial condition overestimate the results. Note that the MV$^\gamma$ and MV$^e$ models are calibrated by fitting the DIS data from HERA; the agreement with the LHC data suggests that the CGC framework can capture the essential small-$x$ QCD dynamics across different collision energies and systems. %While the rcBK dipole amplitude is flavor-independent, we find our calculation underestimates the $\eta'$ production, which indicates that identified particles are dominantly determined by the fragmentation processes.

%\begin{figure*}[tbh]
%    \centering
%        \includegraphics[width=0.42\textwidth]{Fig_dhjkt/dnchdetapp7.pdf}
%    \caption{$\d N_{ch}\d\eta$ measured in pp 7 TeV simulated with DHJ and $k_T$ factorization separately. Experimental data taken from~\cite{LHCb:2014wmv}. \blue{In the lower panel, we present the ratio between our simulation and the experimental measurements.} \com{\bf why are there two groups in the right panel?}\label{fig:dhjkt1}}
%\end{figure*}

\subsubsection{Particle production at RHIC and at mid-rapidity at LHC}
\begin{figure*}[tbh]
    \centering
        \includegraphics[width=0.42\textwidth]{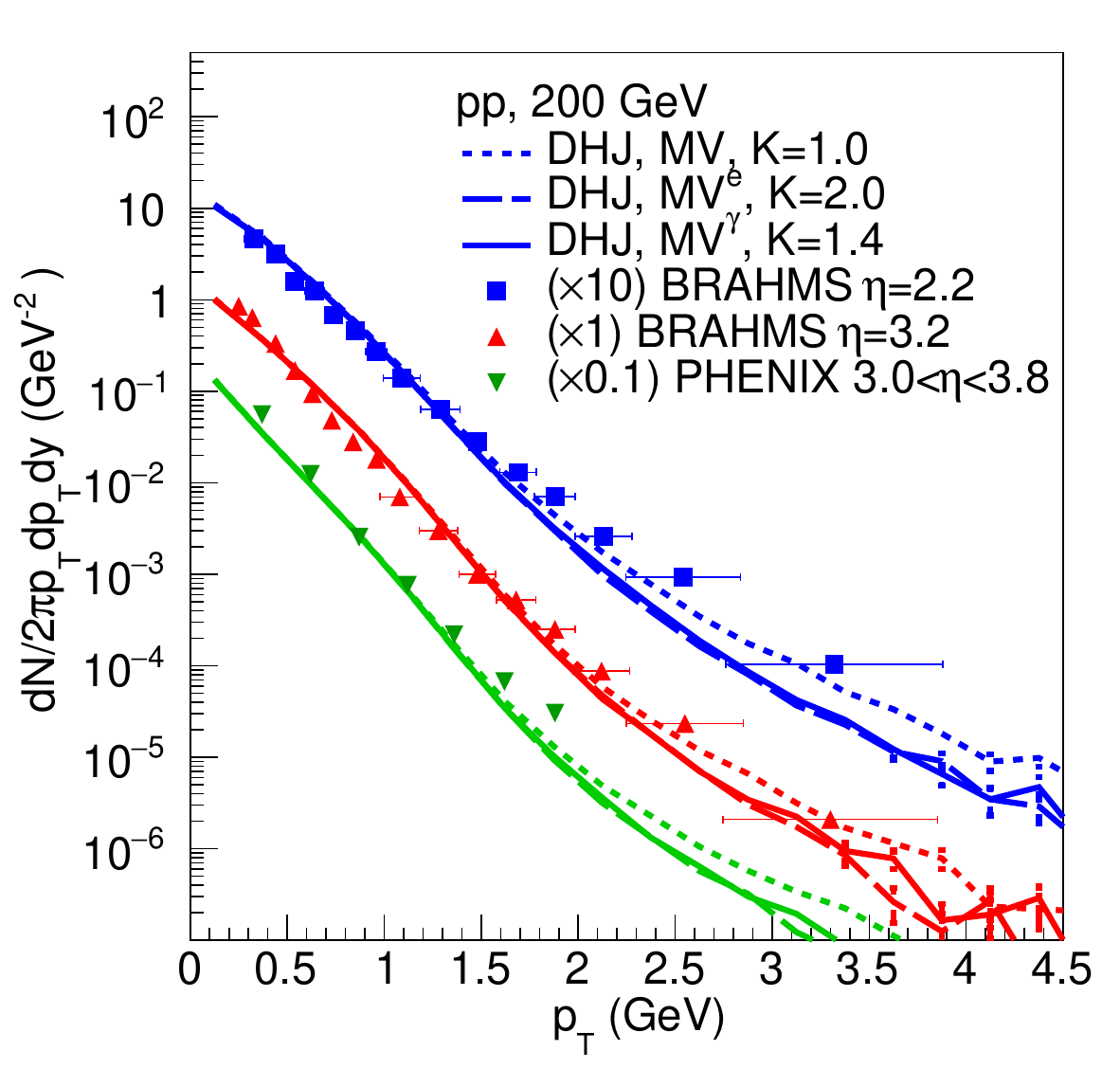}
    \hspace{0.5cm}
        \includegraphics[width=0.42\textwidth]{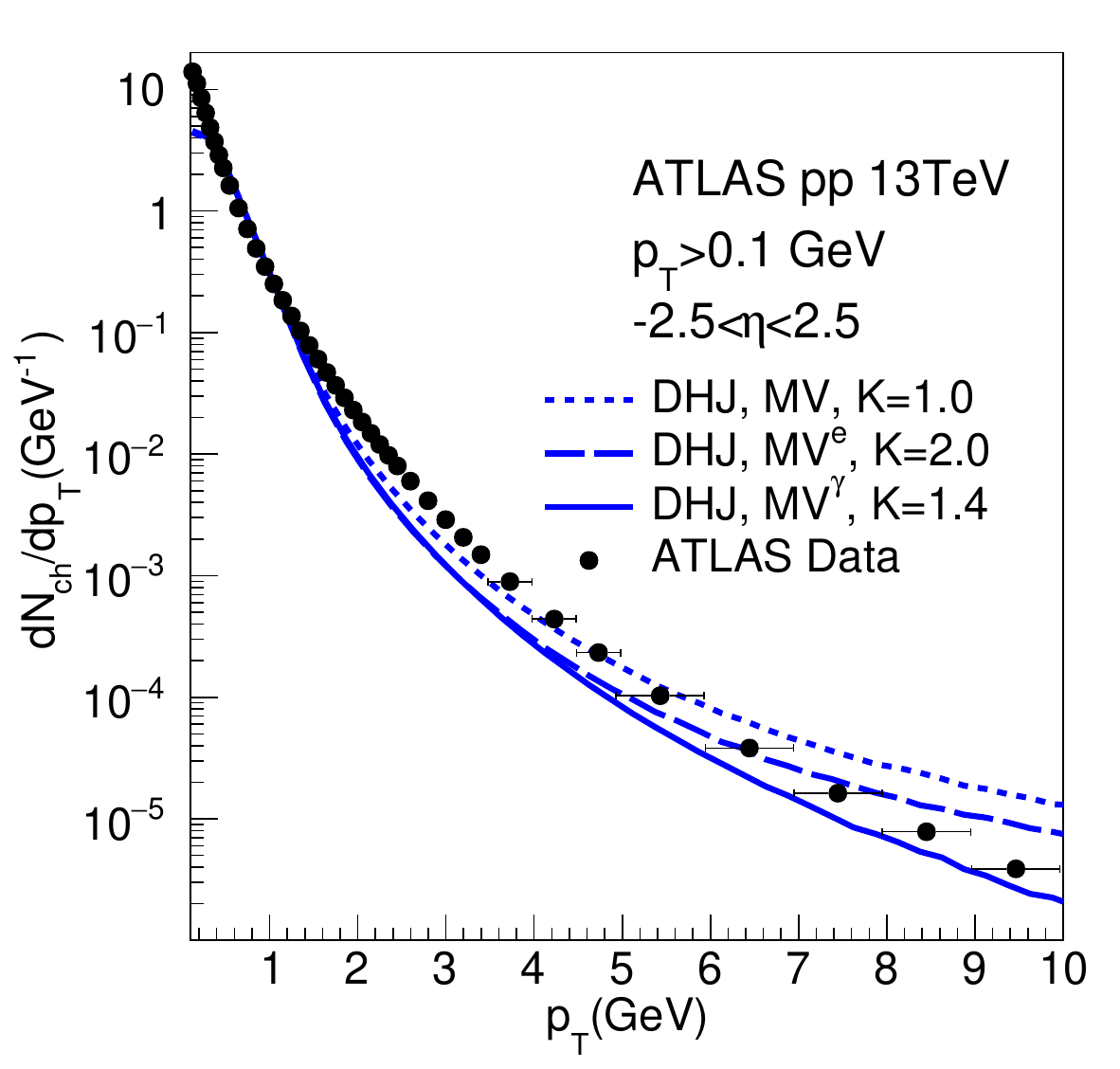}
    \caption{$p_T$ spectra in pp collisions at $\sqrt{s}=200\rm\ GeV$ at RHIC (left) and in pp collisions at $\sqrt{s}=13\rm\ TeV$ in mid-rapidity at LHC (right). The BRAHMS data is measured with negatively charged particles taken from Ref.~\cite{BRAHMS:2004xry}, the PHENIX data is measured with $\pi^0$ taken from Ref.~\cite{PHENIX:2011puq}, and the ATLAS data is measured with charged hadrons taken from Ref.~\cite{ATLAS:2016zba}. Model simulations are performed based on the DHJ formula with the three different initial conditions and corresponding normalization factor $K$: MV model (dotted lines), MV$^e$ model (dashed lines), and MV$^\gamma$ model (solid lines).}\label{fig:fig7}
\end{figure*}

To extend the evaluation across different kinematical regions, Fig.~\ref{fig:fig7} presents the $p_T$ spectra at RHIC collision energy of $\sqrt{s}=200$ GeV (left panel) and at the mid-rapidity region at the LHC energy of $\sqrt{s} = 13$ TeV (right panel). Despite several minor modifications relative to the earlier implementation~\cite{Deng:2014vda}, the current model maintains a good description of the RHIC data. At RHIC energies, the typical momentum fraction $x$ is $x \gtrsim 10^{-1}$ for the dilute projectile and $x \lesssim 10^{-2}$ for the dense target, which marginally supports the applicability of the DHJ model.  
In contrast, the DHJ model shows a clear discrepancy with the measured data at mid-rapidity at the LHC energies, as shown in the right panel of Fig.~\ref{fig:fig7}. 
Increasing the $K$ factor in the DHJ calculation with the MV$^\gamma$ initial condition improves the agreement with the $p_T$ spectrum of the data, but leads to an overestimate of the particle yield at forward rapidity.
%By increasing the $K$ factor, the DHJ calculation with the MV$^\gamma$ initial condition can reproduce the data accordingly, but with the overall overestimate of the particle yield in forward rapidity.
%\YN{In ref.[60] fig 3 and 4, it seems that DHJ describes the data well at 1.96 7 TeV. Does the difference come from the fragmentation function or from energy?}

\begin{figure*}[tbh]
    \centering
        \includegraphics[width=0.42\textwidth]{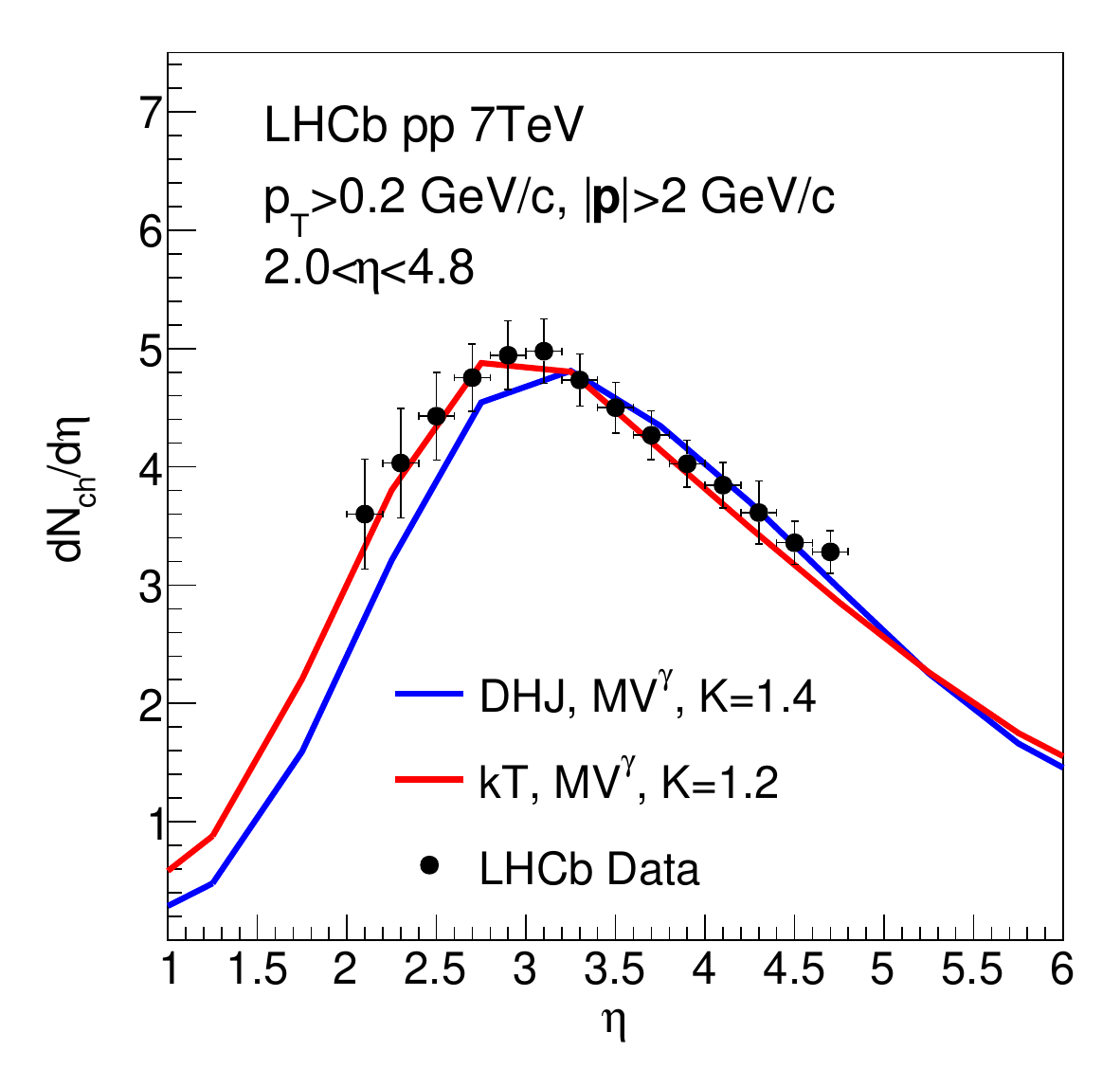}
    \hspace{0.5cm}
        \includegraphics[width=0.42\textwidth]{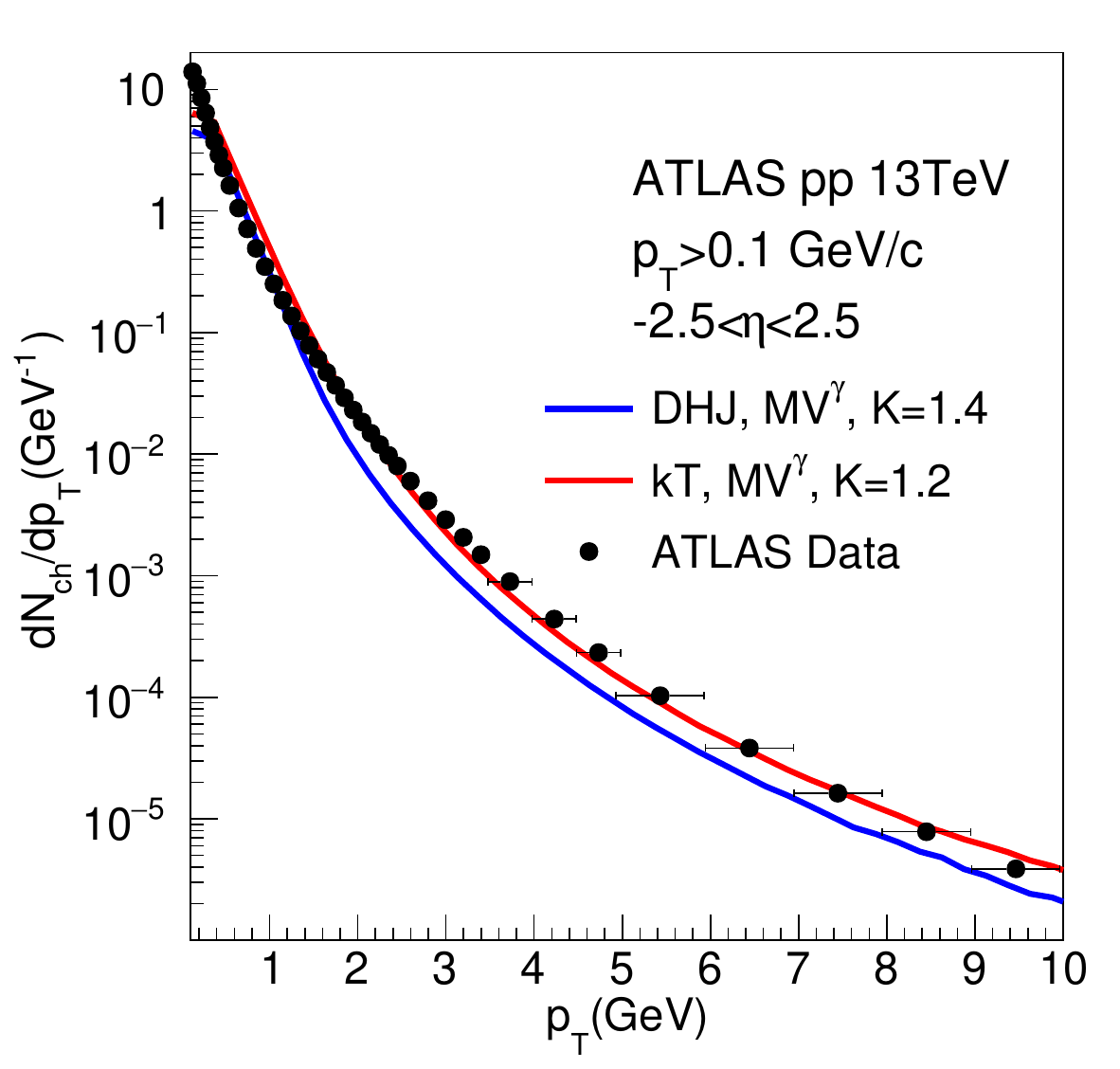}
    \caption{Comparison between the DHJ and $k_T$ factorization frameworks for the charged particle pseudo-rapidity distribution $\d N_{\rm ch}/\d\eta$ (left) and  $p_T$ spectra $\d N_{\rm ch}/\d p_T$ (right) in pp collisions at $\sqrt{s}=7$ TeV. Experimental data taken from Refs.~\cite{LHCb:2014wmv, ATLAS:2016zba}. Model simulations are performed based on the DHJ formula (blue) and the $k_T$-factorization formula (red) with the same MV$^\gamma$ initial condition.}\label{fig:fig8}
\end{figure*}

As discussed in Fig.~\ref{fig:xrange}, mid-rapidity pp collisions at the LHC energies are more appropriately described within the $k_T$-factorization framework.
This motivates the calculation shown in Fig.~\ref{fig:fig8}, where we present the comparison of hadron production from the DHJ and $k_T$-factorization approaches, using the same MV$^\gamma$ parameterization. For consistency, the normalization factor $K$ of the $k_T$ factorization is fixed by matching the forward rapidity particle production from the DHJ model; predictions are then made for the mid-rapidity region. Within this setup, the $k_T$ factorization framework provides a more reasonable description than the DHJ model in the mid-rapidity region at the LHC energies. In particular, it represents the correct shape of $p_T$ spectra, which is consistent with the expectation that mid-rapidity pp collisions at the LHC energies are significantly influenced by dense gluon dynamics from both projectile and target. However, more observables should be investigated before drawing a solid conclusion.

As an additional check, we compared the results of the DHJ calculation with those from a collinear-factorization-based \textsc{PYTHIA} simulation at both RHIC and LHC energies~\cite{LHCb:2023iyw, LHCb:2021abm, Aguilar:2021sfa}. The inclusive hard-$p_T$ spectra show broadly similar behavior in the two approaches, suggesting that these observables alone have limited sensitivity to the underlying small-$x$ dynamics. More noticeable differences are found in the soft-production region~\cite{LHCb:2014wmv}. However, these differences cannot be uniquely attributed to saturation effects because the soft component is also sensitive to the phenomenological treatment of parton showering, hadronization, and the underlying event. This comparison therefore provides a useful collinear-factorization baseline, while more differential observables are required to assess saturation and low-$x$ evolution effects more directly.

\subsection{Predictions for Future FoCal Measurements}\label{sec:predicsion-for-future-focal-measurements}
\subsubsection{Hadron production}

\begin{figure*}[tbh]
    \centering
        \includegraphics[width=0.42\textwidth]{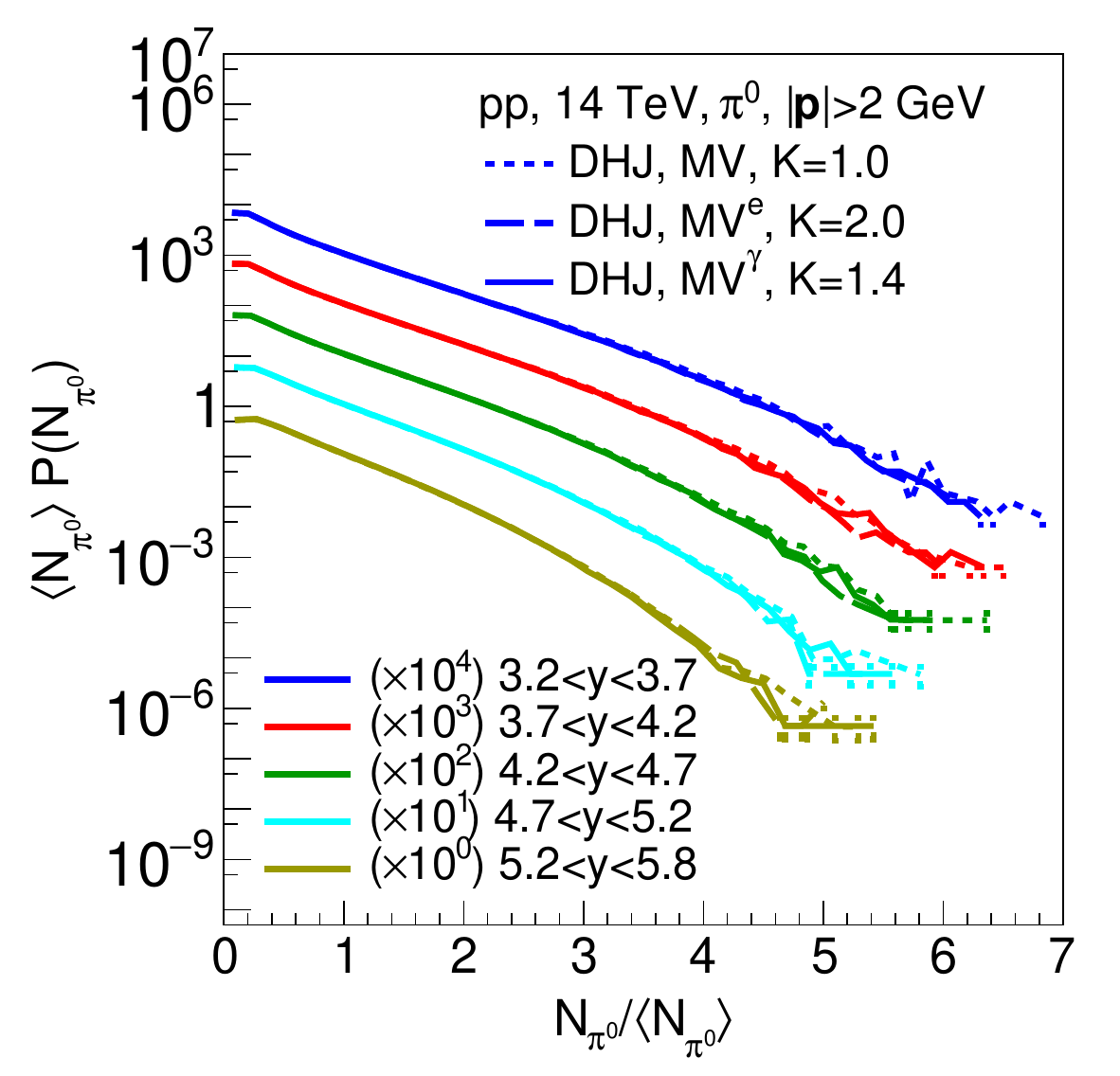}
    \hspace{0.5cm}
        \includegraphics[width=0.42\textwidth]{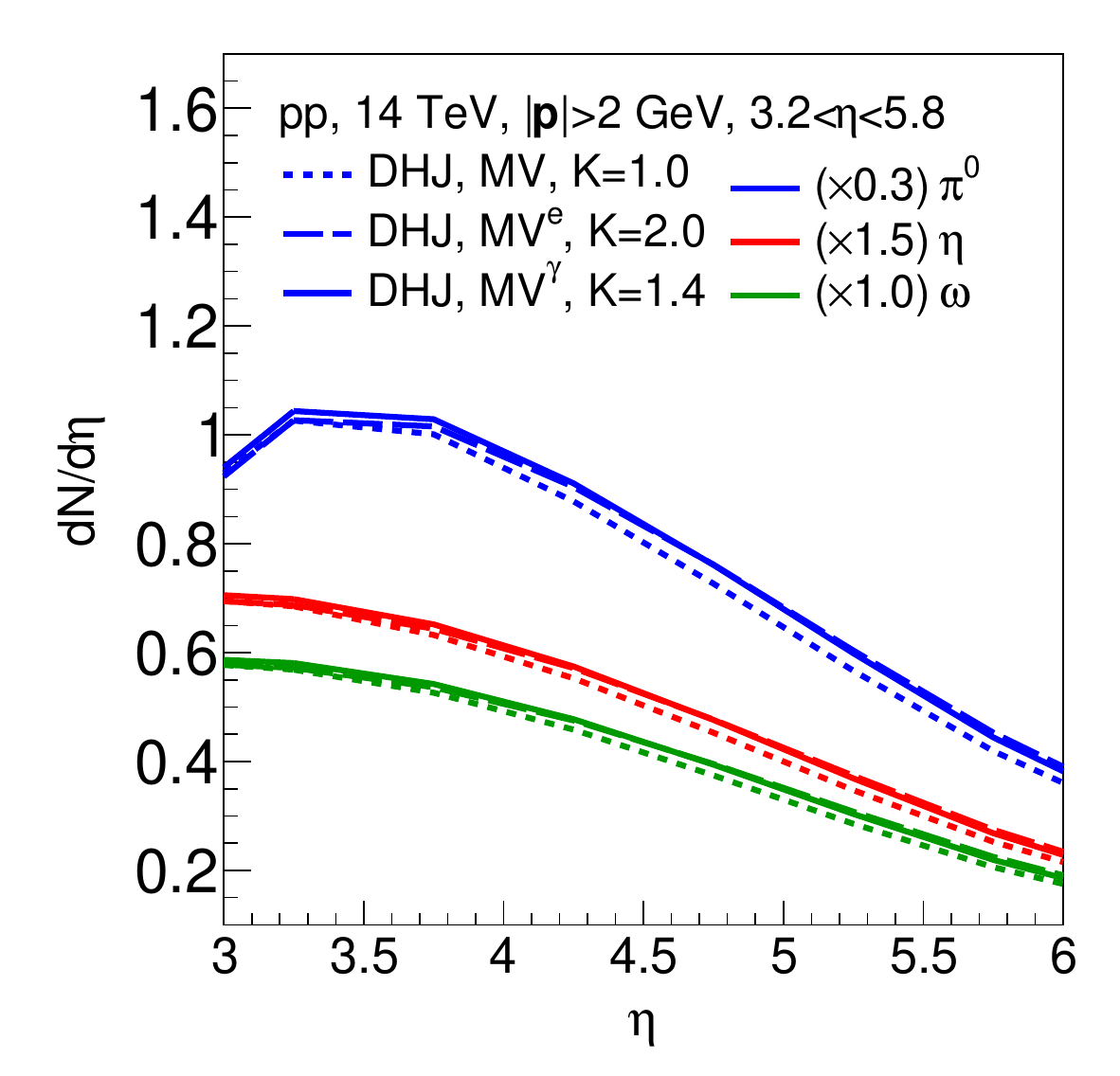}
    \caption{Neutral pion multiplicity distribution (left) and neutral particle ($\pi^0$, $\eta$, and $\omega$) pseudo-rapidity distribution $\d N/\d\eta$ (right) with $0.0<p_T<12\rm\ GeV$ in pp collisions at $\sqrt{s}=14\rm\ TeV$. Model simulations are performed based on the DHJ formula with the three different initial conditions and corresponding normalization factor $K$: MV model (dotted lines), MV$^e$ model (dashed lines), and MV$^\gamma$ model (solid lines).} \label{fig:fig9}
\end{figure*}

In this section, we present predictions for future measurements with the FoCal detector. The left panel of Fig.~\ref{fig:fig9} shows the predicted normalized multiplicity distribution of neutral pions, $\langle N_{\pi^0}\rangle P(N_{\pi^0})$ in pp collisions at $\sqrt{s}=14$ TeV. With increasing rapidity, the normalized distribution narrows, indicating reduced multiplicity fluctuations in the forward region. %\Remove{We note, however, that our current model does not include event-by-event fluctuations of the saturation scale $Q_s^2$, and a further study incorporating such fluctuations is left for future work. }
%\YN{If we comment on the fluctuations of the saturation scale, do you expect its effects to be large?}
The right panel of Fig.~\ref{fig:fig9} displays the predicted pseudo-rapidity density $\d N/\d\eta$ for neutral particles with the cut $|\bm{p}| >2 {\rm\ GeV}$. As expected, the three parameterizations present a similar shape, with the MV model slightly different from the other two.

\begin{figure*}[tbh]
    \centering
        \includegraphics[width=0.42\textwidth]{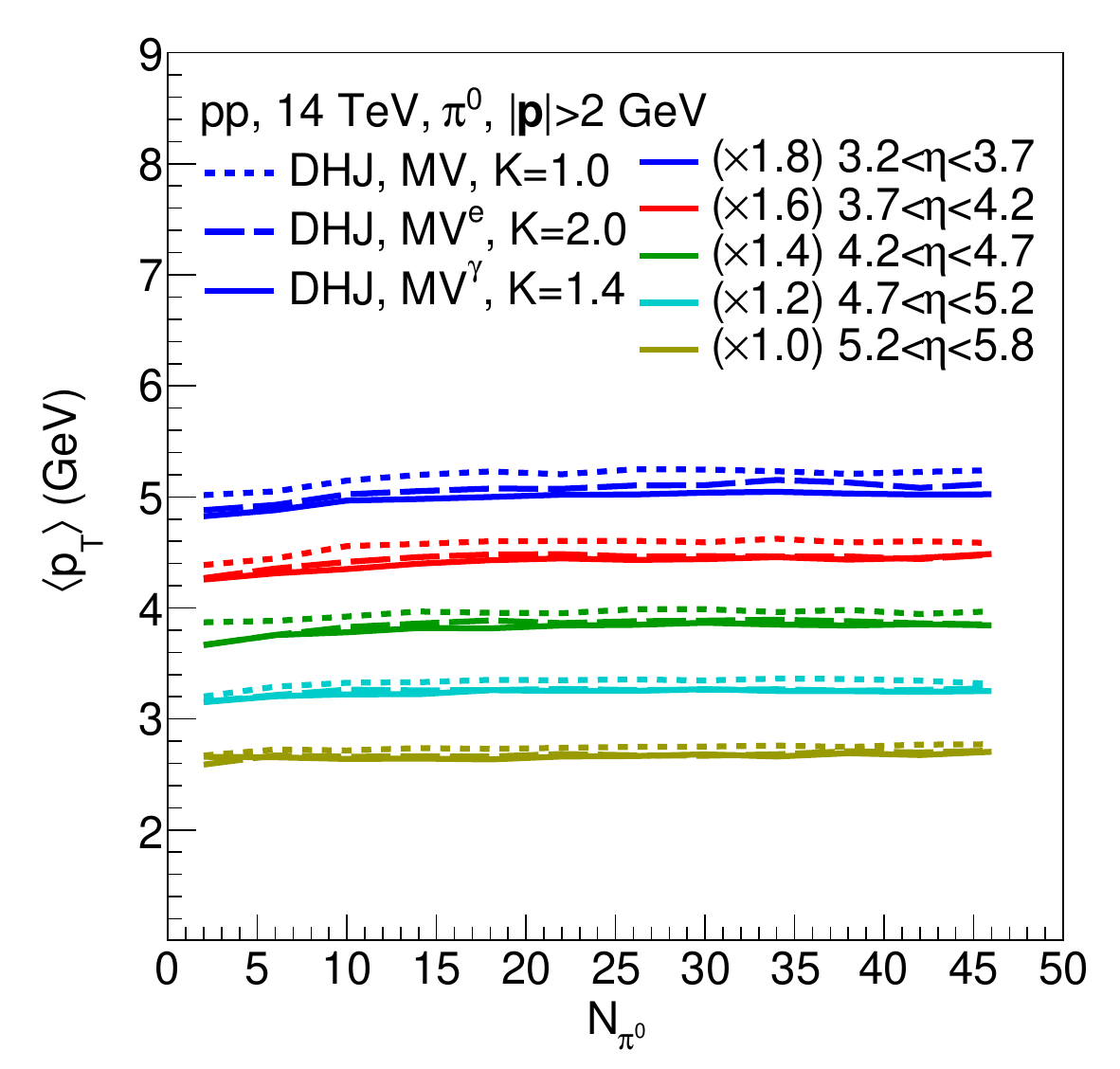}
    \hspace{0.5cm}
        \includegraphics[width=0.42\textwidth]{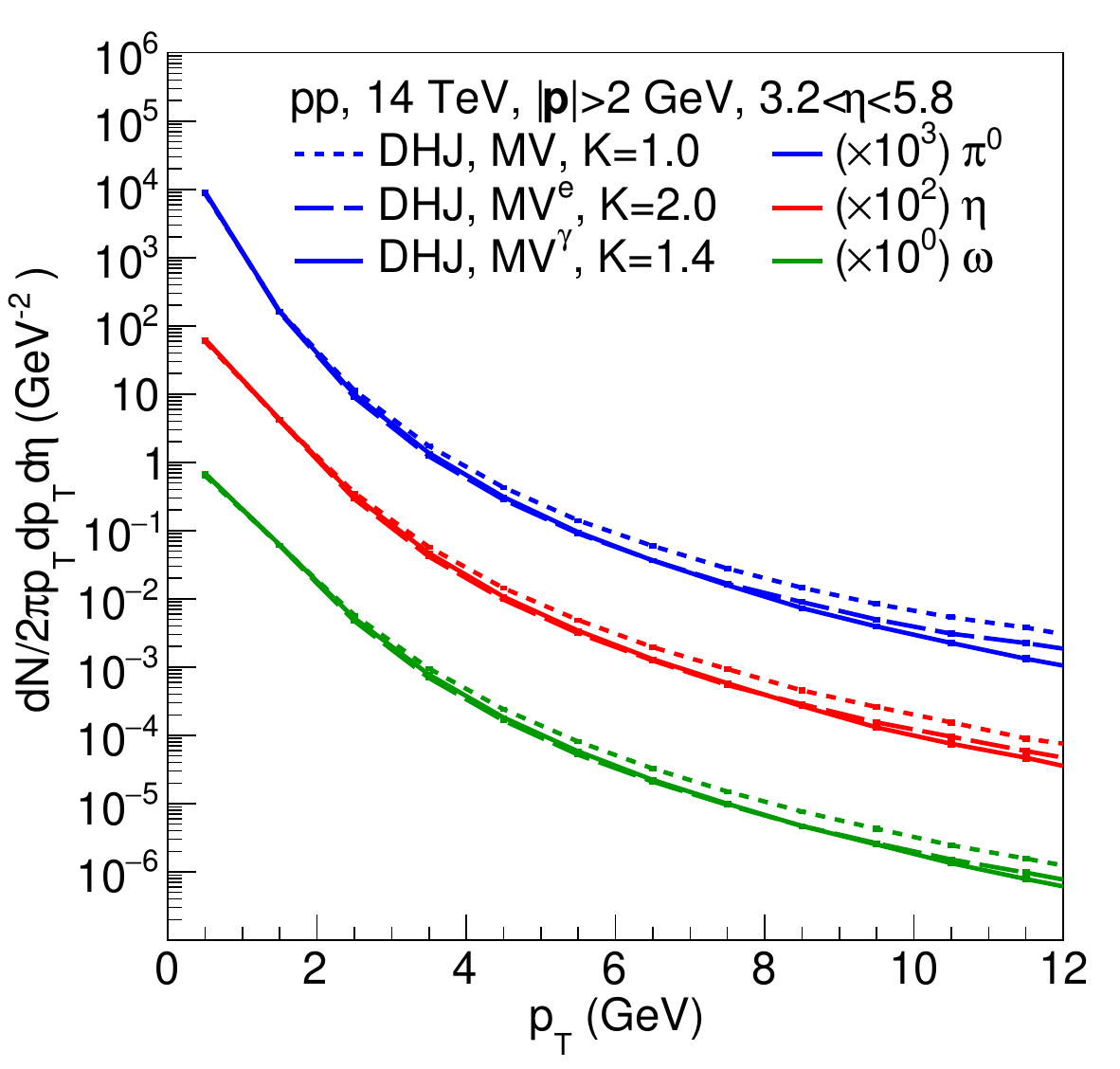}
    \caption{Neutral pion mean transverse momentum (left) and neutral particle ($\pi^0$, $\eta$, $\omega$) $p_T$ spectra (right) predicted in pp collisions at $\sqrt{s}=14\rm\ TeV$ with $3.2<\eta<5.8$. Model simulations are performed based on the DHJ formula with the three different initial conditions and corresponding normalization factor $K$: MV model (dotted lines), MV$^e$ model (dashed lines), and MV$^\gamma$ model (solid lines).}  \label{fig:fig10}
\end{figure*}

Figure~\ref{fig:fig10} shows the neutral-pion multiplicity $N_{\pi^0}$ dependent mean transverse momentum $\langle p_T\rangle$ in the left panel, and the $p_T$ spectra for $\pi^0$, $\eta$, and $\omega$ mesons, with the full FoCal pseudo-rapidity acceptance, in the right panel.
%In the left panel of Fig.~\ref{fig:fig10}, we present the neutral-pion multiplicity $N_{\pi^0}$ dependent mean transverse momentum $\langle p_T\rangle$. In the right panel of Fig.~\ref{fig:fig10}, we present the $p_T$ spectra for $\pi^0$, $\eta$, and $\omega$ mesons, with the full FoCal pseudo-rapidity acceptance. 
The predictions exhibit a stronger dependence on the initial condition at harder $p_T>6\rm\ GeV$. While the $x$ value on the target side is about $x_t\sim10^{-5}$ in this region, the small proton saturation scale leads to the distinction between different initial conditions. Precise measurements in this regime could help constrain the shape of the dipole amplitude in the initial condition, and simultaneously provide valuable insight into the kinematic threshold below which the DHJ formalism remains valid.

\begin{figure*}[tbh]
    \centering
        \includegraphics[width=0.32\textwidth]{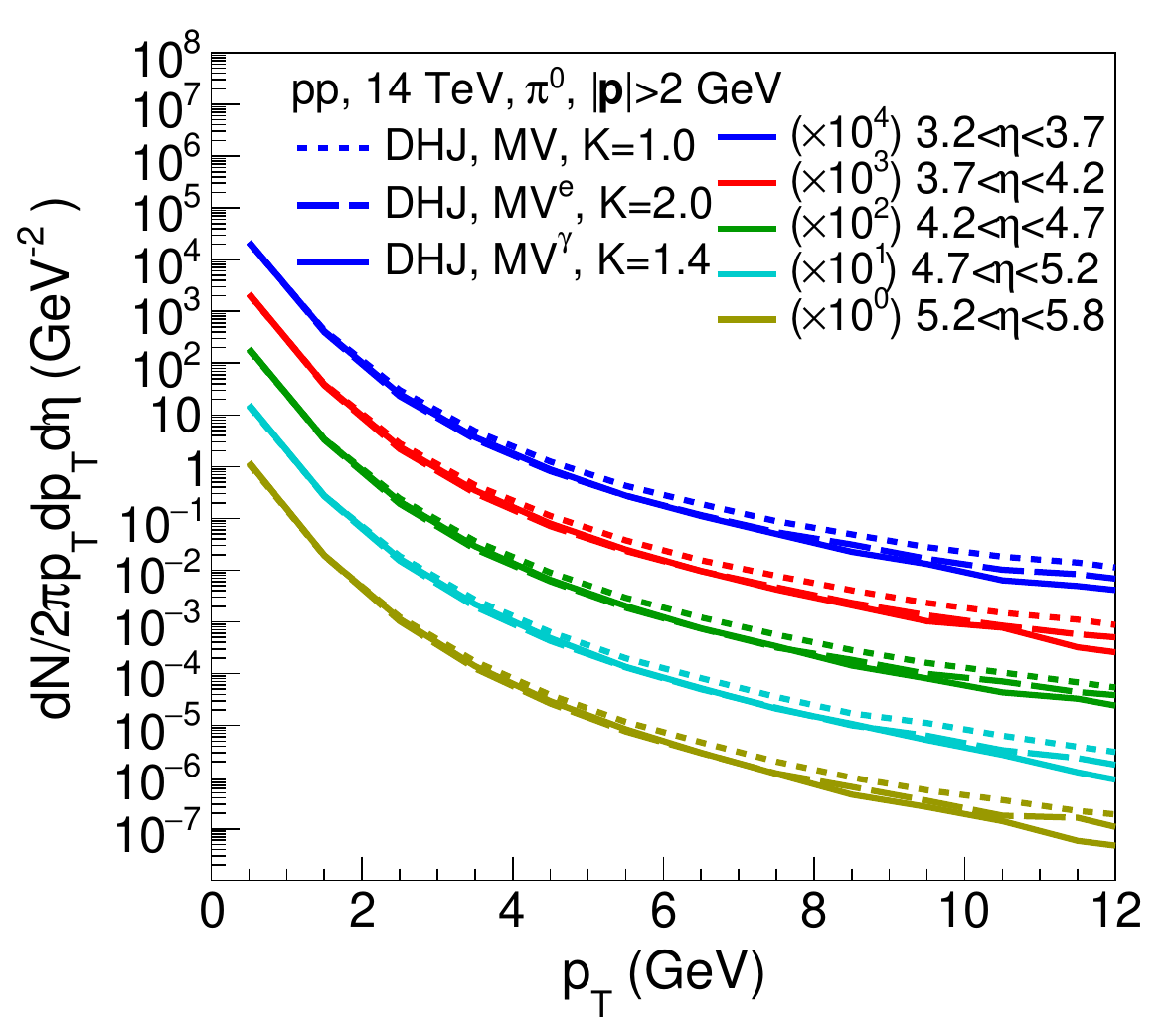}
    \hspace{0.1cm}
        \includegraphics[width=0.32\textwidth]{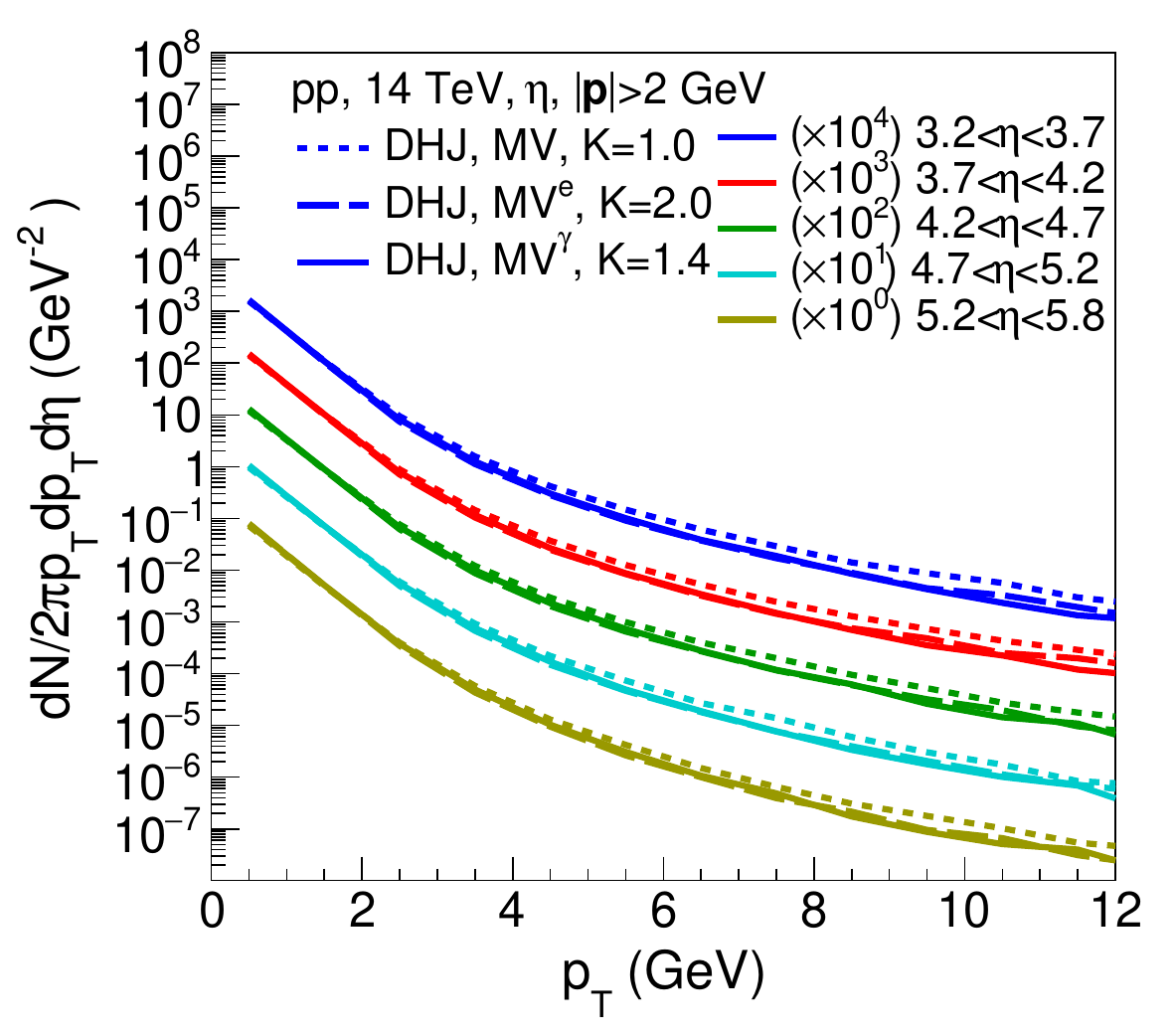}
    \hspace{0.1cm}
        \includegraphics[width=0.32\textwidth]{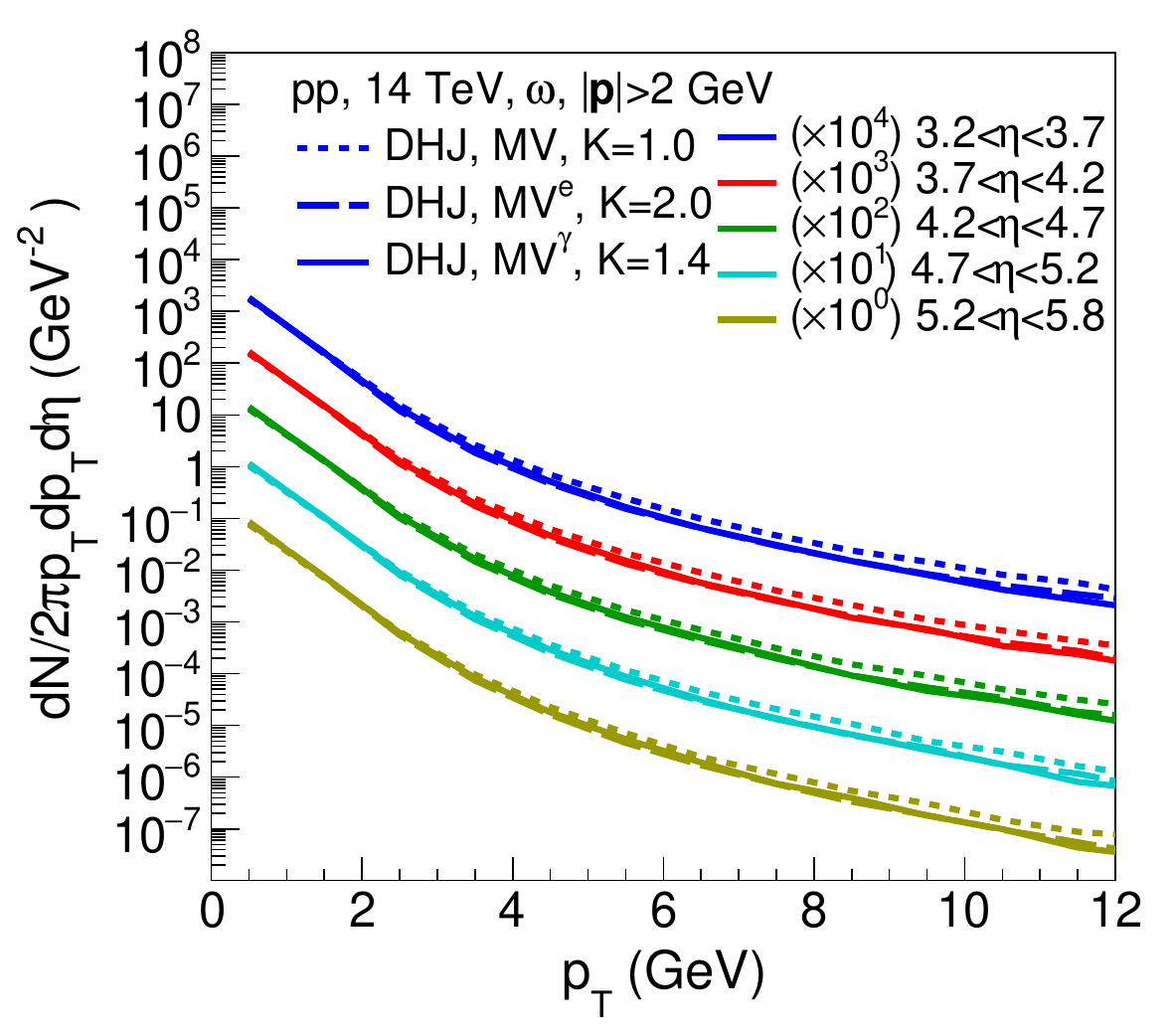}
    \caption{$p_T$ spectra of  $\pi^0$ (left),  $\eta$ (middle), and  $\omega$ meson (right) simulated in pp collisions at $\sqrt{s}=14\rm\ TeV$. Model simulations are performed based on the DHJ formula with the three different initial conditions and corresponding normalization factor $K$: MV model (dotted lines), MV$^e$ model (dashed lines), and MV$^\gamma$ model (solid lines).} \label{fig:fig11}
\end{figure*}

In Fig.~\ref{fig:fig11}, we present the predicted $p_T$ spectra for different particle species across several pseudo-rapidity bins. Within the intermediate $p_T$ range ($2\mathrm{\ GeV}<p_T<6\mathrm{\ GeV}$), the MV$^e$ and MV$^\gamma$ models yield similar spectral shapes, both exhibiting steeper slopes than the MV model. At very high $p_T$ ($p_T>10\rm\ GeV$), the MV and MV$^e$ models produce flatter spectra compared to MV$^\gamma$. %\Remove{This difference stems primarily from the $\gamma$-dependence of the dipole amplitude, which governs the high-$p_T$ tail scaling%~\cite{Dumitru:2005gt, Dumitru:2005kb, Hayashigaki:2006ek}. The larger value of $\gamma=1.101$ in the MV$^\gamma$ model thus leads to a steeper power-law fall-off.} 
Moreover, the sensitivity to the initial condition depends noticeably on the particle species. The $\pi^0$ spectrum displays the strongest model dependence, while the effects are reduced for $\eta$ and $\omega$ mesons. This variation may arise from the interplay between the CGC-based scattering dynamics, in which the projectile PDFs determine the parton-flavor component during parton generation, and the subsequent \textsc{PYTHIA} string-fragmentation process that determines the final-state hadron species.

\begin{figure*}[tbh]
    \centering
        \includegraphics[width=0.32\textwidth]{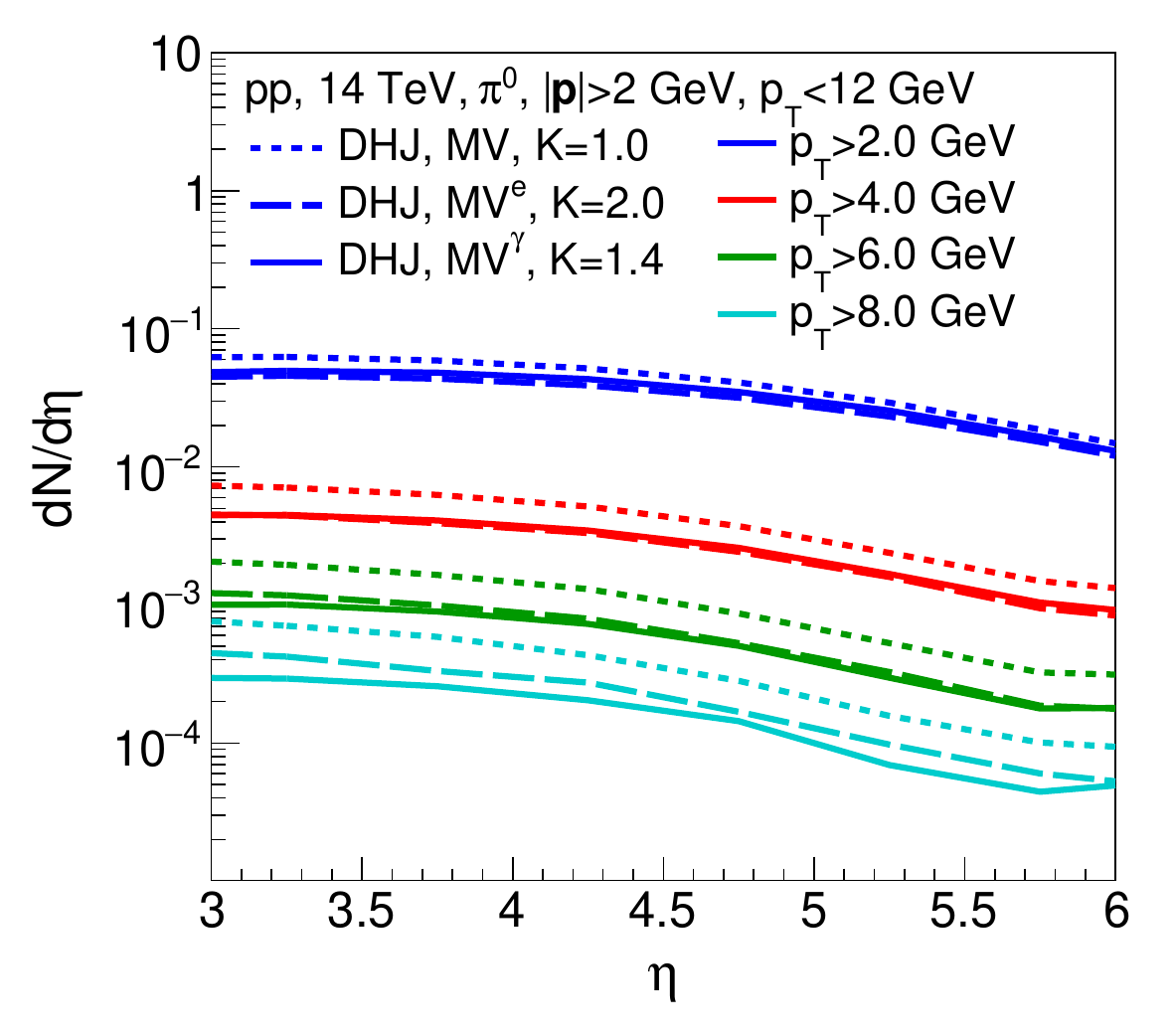}
    \hspace{0.1cm}
        \includegraphics[width=0.32\textwidth]{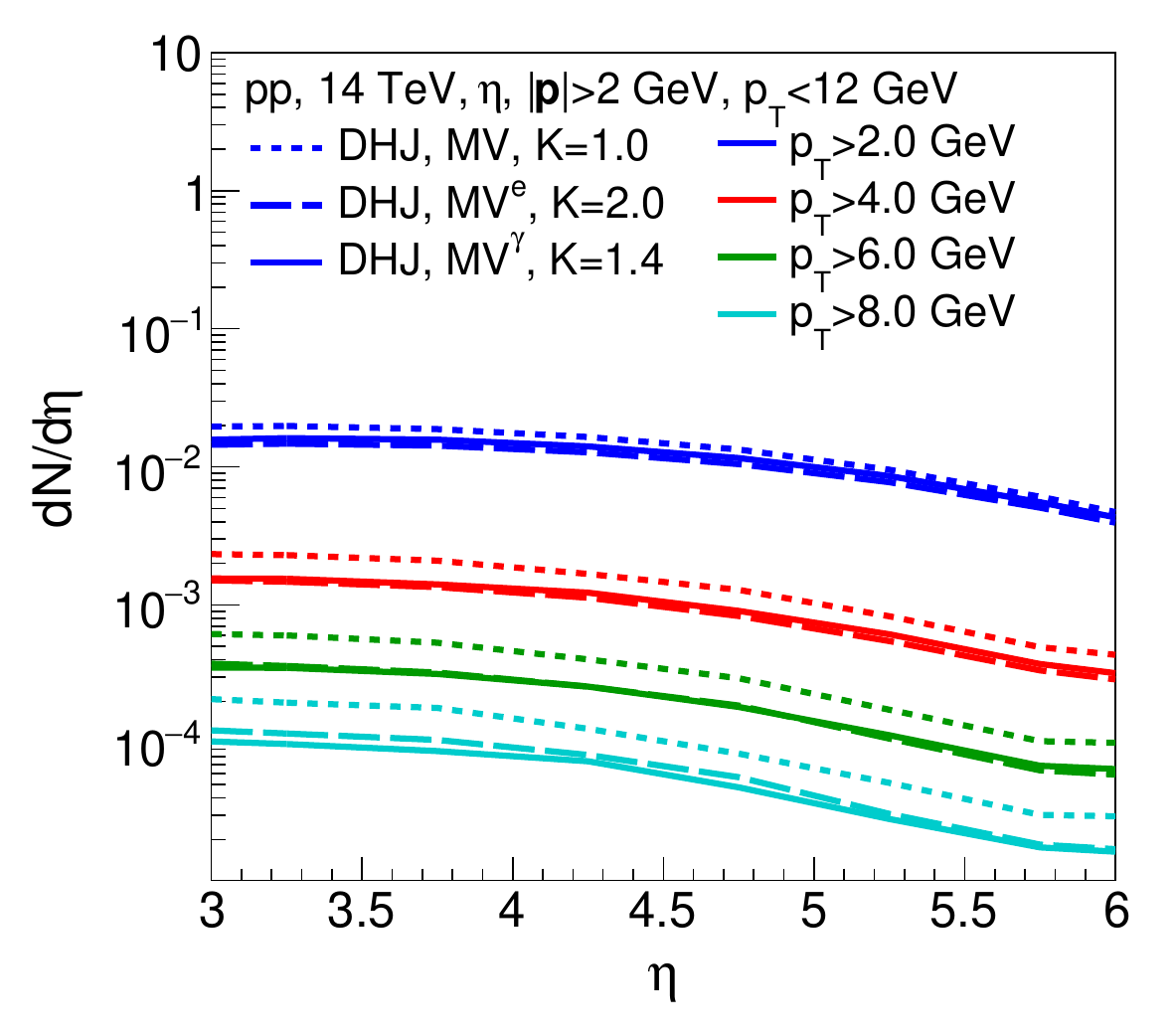}
    \hspace{0.1cm}
        \includegraphics[width=0.32\textwidth]{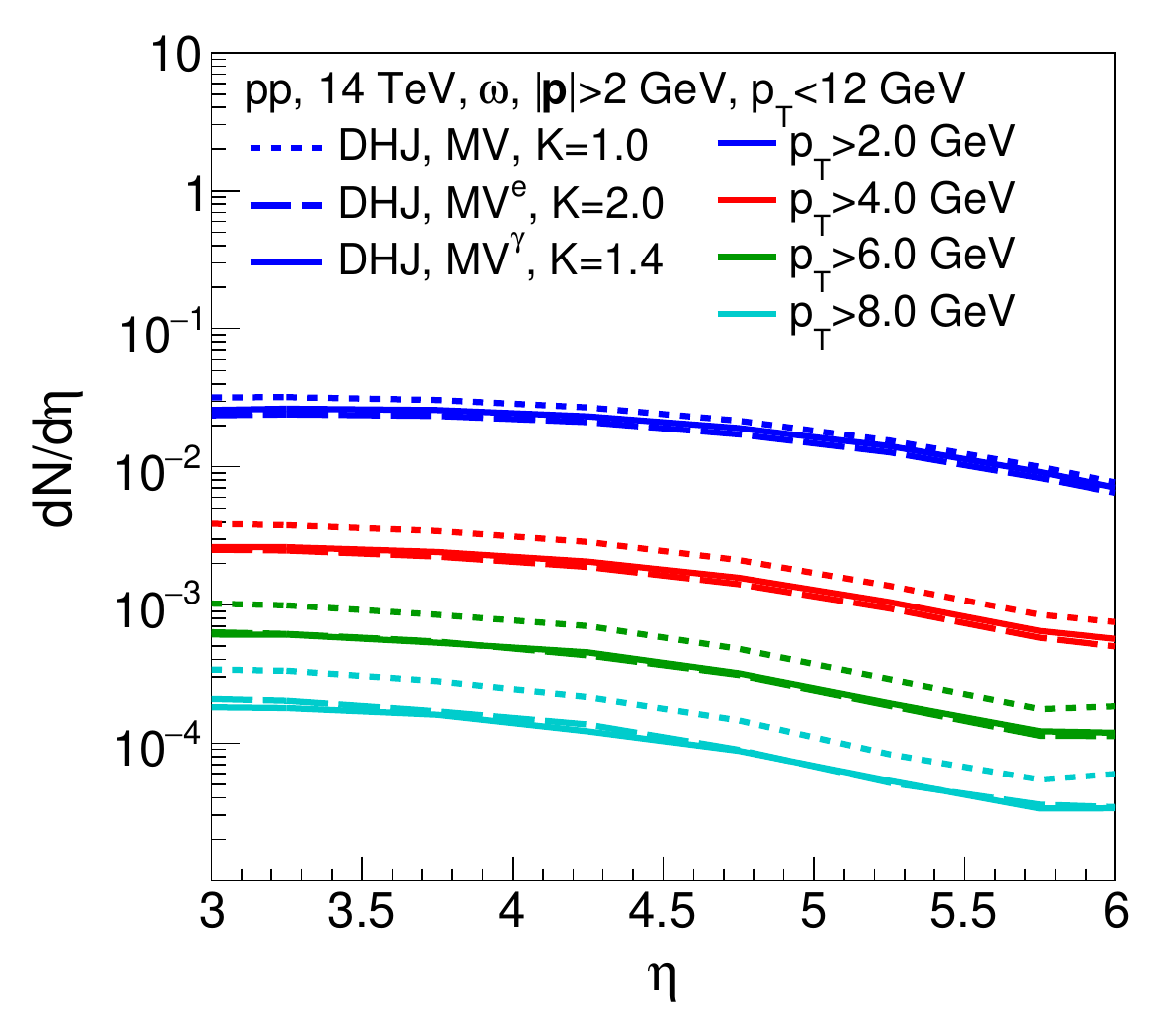}
    \caption{$\d N/\d\eta$ distribution of  $\pi^0$ (left), $\eta$ (middle), and $\omega$ mesons (right) simulated in pp collisions at $\sqrt{s}=14\rm\ TeV$. Model simulations are performed based on the DHJ formula with three different initial conditions and corresponding normalization factors $K$: MV model (dotted lines), MV$^e$ model (dashed lines), and MV$^\gamma$ model (solid lines).} \label{fig:fig12}
\end{figure*}

In Fig.~\ref{fig:fig12}, we present the pseudo-rapidity density $\d N/\d\eta$ for neutral particles under various $p_T$ thresholds. The spectra with a low $p_T$ cut show less sensitivity to the choice of rcBK initial condition. 
In contrast, sensitivity can be observed with a higher $p_T$ cut. 
%\Remove{We further observe that the differences between MV$^\gamma$ and MV$^e$ parameterizations diminish at more forward rapidities, reflecting the attractor behavior of the BK evolution.} 
Consequently, measurements of $\d N/\d\eta$ with a high $p_T$ cut in pp collisions offer an opportunity to constrain the parameterizations for the initial dipole amplitude, thus reducing the theoretical uncertainty. The figure also includes predictions for $\eta$ and $\omega$ meson productions, which show a similar trend compared with neutral pion production.

\subsubsection{Jet production}
\begin{figure*}[tbh]
    \centering
        \includegraphics[width=0.32\textwidth]{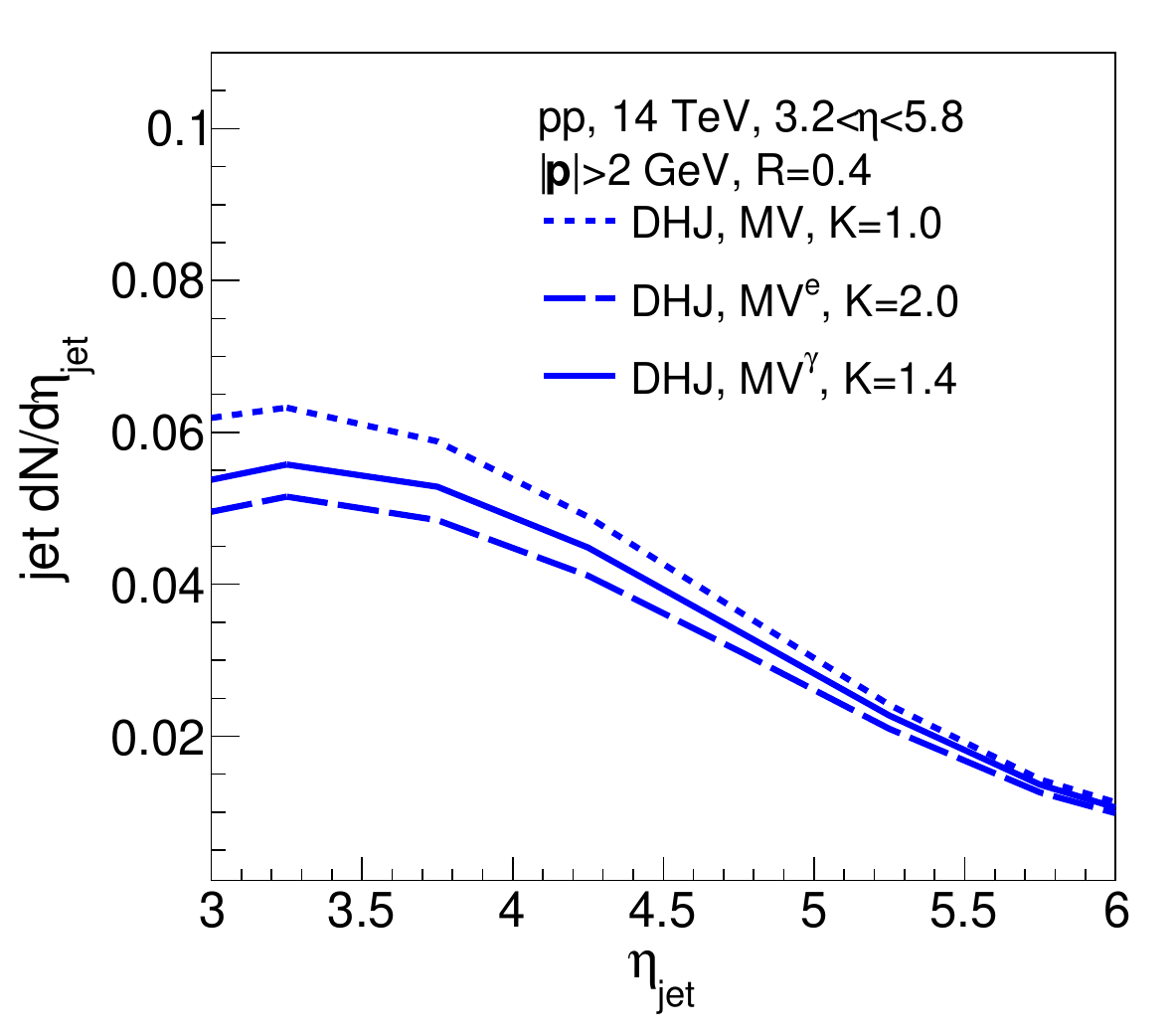}
    \hspace{0.1cm}
        \includegraphics[width=0.32\textwidth]{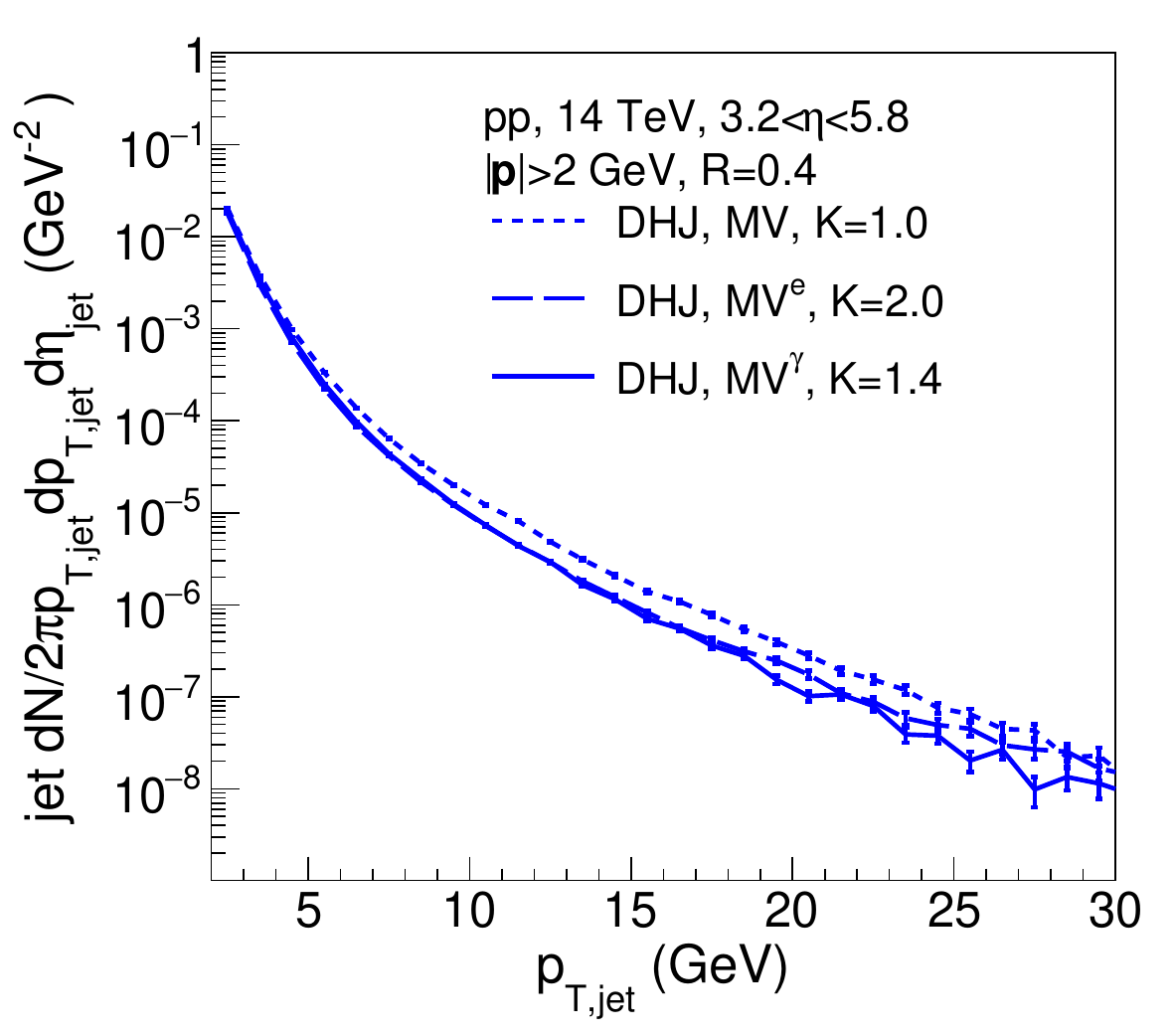}
    \hspace{0.1cm}
        \includegraphics[width=0.32\textwidth]{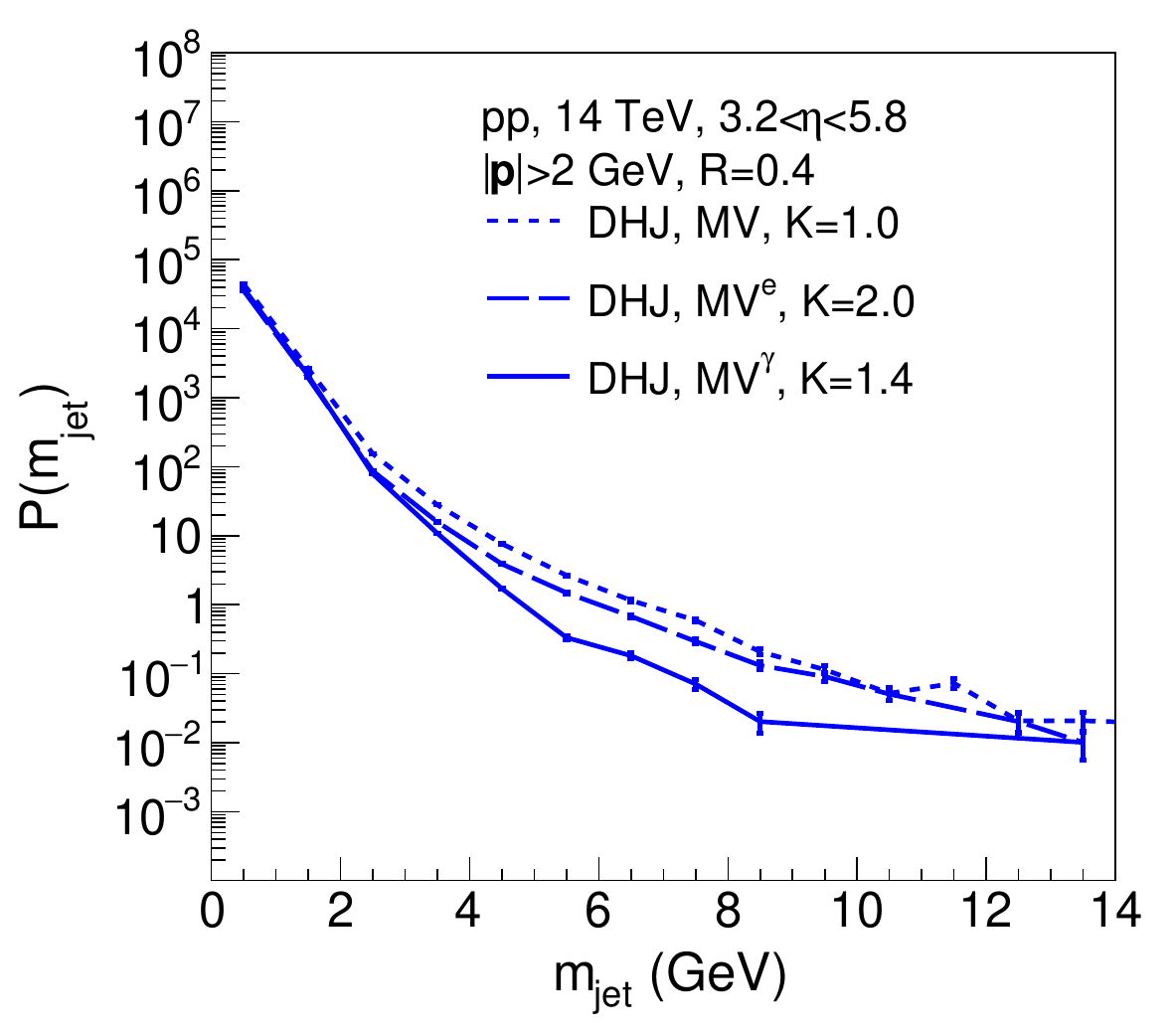}
    \caption{jet $\d N/\d\eta$ (left), jet $\d N/\d p_{T}$ spectra (middle) and  jet mass distribution $P(m_{\rm jet})$ (right) in pp collisions at $\sqrt{s}=14\rm\ TeV$. jets are selected with $p_{T,\rm\ jet}>2\rm\ GeV$ and $p_{T,\rm\ hardest}<12\rm\ GeV$. Model simulations are performed based on the DHJ formula with the three different initial conditions and corresponding normalization factor $K$: MV model (dotted lines), MV$^e$ model (dashed lines), and MV$^\gamma$ model (solid lines).} \label{fig:fig13}
\end{figure*}

In the final part of this section, we provide predictions of jet observables, which are expected to have reduced sensitivity to the fragmentation effects. In these calculations, jets are reconstructed from neutral pions using the anti-$k_T$ algorithm~\cite{Cacciari:2008gp} with a radius parameter $R=0.4$, and a soft transverse-momentum cut $p_{T,\mathrm{jet}}>2\rm\ GeV$ is imposed. To retain the sensitivity to saturation effects, the observables are constructed from reconstructed jets for which the leading (highest-$p_T$) constituent satisfies $p_{T}^{\rm hardest}<12$ GeV, while no direct upper cut is imposed on the reconstructed jet transverse momentum.
%\YN{ pt<12 seems not soft in the sense of CGC because it is much higher than the saturation scale.}
We have verified that the resulting jet spectra are stable under variations of the jet radius within the range $0.2<R<0.6$.

Figure~\ref{fig:fig13} presents the prediction of the jet pseudo-rapidity distribution $dN/d\eta$, transverse-momentum spectra $\d N/\d p_T$, as well as the jet mass distribution $P(m_{\rm jet})$ in pp collisions at $\sqrt{s}=14$ TeV. 

The pseudo-rapidity distribution indicates that the differences among the three rcBK initial conditions are more pronounced at mid‑rapidity and gradually diminish toward forward rapidity. %The behavior is consistent with the fact that the nonlinear small-$x$ evolution will reduce the sensitivity to the initial condition at forward rapidity.
The difference among the models becomes more evident in the jet $p_T$ spectra, shown in the middle panel of Fig.~\ref{fig:fig13}. Deviations between the MV model and the modified versions (MV$^\gamma$ and MV$^e$ models) first appear in the jet $p_T\sim 5\rm\ GeV$ region. At higher jet momenta, an ordering appears, with the MV model predicting the largest yield, followed by the MV$^e$ and MV$^\gamma$ models.
%\red{The behavior of the jet $\eta$ and jet $p_T$ spectra reflects the sensitivity of these observables to different kinematic regions of the dipole amplitude. The jet $\eta$ spectra shown in the left panel of Fig.~\ref{fig:fig13} receive relatively larger contributions from soft-$p_T$ jet production and are therefore more sensitive to the treatment of the softer-$k_T$ region of the dipole amplitude. In contrast, the jet $p_T$ spectra shown in the middle panel provide a more direct probe of the larger transverse-momentum region. Varying $R$ within the range $0.2 \le R \le 0.8$ or imposing a smaller $p_T$ cut on the leading hadron does not qualitatively change the results, suggesting that the observed differences mainly originate from the underlying hard-scattering dynamics. Therefore, the combined study of these two jet observables provides complementary information that can help constrain the shape of the dipole amplitude more precisely.}

The behavior of the jet $\eta$ and jet $p_T$ spectra reflects the sensitivity of these observables to different kinematic regions of the dipole amplitude. The jet $\eta$ spectra shown in the left panel of Fig.~\ref{fig:fig13} receive relatively larger contributions from soft-$p_T$ jet production and are therefore more sensitive to the treatment of the softer-$k_T$ region of the dipole amplitude. In contrast, the jet $p_T$ spectra shown in the middle panel provide a more direct probe of the larger transverse-momentum region. Varying $R$ within the range $0.2 \le R \le 0.8$ or imposing a smaller $p_T$ cut on the leading hadron does not qualitatively change the results, suggesting that the observed differences mainly originate from the underlying hard-scattering dynamics. Therefore, the combined study of these two jet observables provides complementary information that can help constrain the shape of the dipole amplitude more precisely.

The right panel of Fig.~\ref{fig:fig13} presents the jet mass distributions $P(m_{\rm jet})$, which also exhibit strong sensitivity to the initial conditions. Specifically, the MV$^\gamma$ model yields a narrower distribution than the MV and MV$^e$ models, with the latter two largely consistent with each other. As we have verified, the distribution of normalized jet mass square $\rho\equiv m_{\rm jet}^2/(p_{T,\rm jet}^2 R^2)$ shows no sensitivity to different parameterizations. Consequently, the observed sensitivity here does not originate from the difference of jet substructure, but is driven by the underlying $p_T$ spectra.

%\blue{Explanation about fig13 (left), different pT, hardest test needed}
%\YN{Can we add some guesses about why some observables are sensitive to the initial condition and some are not?
%}
\section{Conclusion}
\label{sec:conclusion}

In this work, we have investigated several observables measured in RHIC and the LHC using the CGC-inspired Monte Carlo model MC-CGC, and provided predictions for future FoCal measurements. Within the default framework employing the DHJ factorization with the MV$^\gamma$-type rcBK initial condition, the model provides a good description of particle yields in existing data, including multiplicity distributions $P(N_{\rm ch})$, $\d N/\d\eta$ measured in forward LHC, as well as $p_T$ spectra measured in forward RHIC and LHC. The present results demonstrate that the Monte Carlo simulation based on the CGC framework provides a consistent description of several observables across a wide range of collision energy, kinematical region, and rapidity.

We have examined the initial conditions of the rcBK evolution by comparing the MV, MV$^e$, and MV$^\gamma$ parameterizations. 
Our results suggest that current forward LHC measurements favor the MV$^\gamma$ and MV$^e$ parameterizations, whereas the MV initial condition produces a flatter $p_T$ spectrum, which is consistent with the observation
in Ref.~\cite{Albacete:2012xq}. As shown in our analysis, the differences arising from the choice of the initial condition in the rcBK equation are most pronounced at higher $p_T$, where the small-$x$ evolution effects are reduced. We also compared two different theoretical approaches: the dilute--dense framework (DHJ factorization) and the dense--dense framework ($k_T$ factorization). They predict different shapes of $p_T$ spectra in the mid-rapidity region ($\eta<2.5$) at the LHC energy, where the measurements favor the $k_T$ factorization approach, suggesting the possible creation of a dense saturated gluonic state for both projectile and target protons at the LHC energies. 

Predictions for future FoCal measurements have also been provided, including hadron and jet observables. For the hadron observables, we calculated the multiplicity distributions, multiplicity dependence of $\langle p_T\rangle$, as well as transverse momentum $p_T$ and pseudo-rapidity $\eta$ distribution of neutral particles ($\pi^0$, $\eta$, and $\omega$), with which the soft physics can be calibrated, and the saturation effects can be investigated. For the jet observables, the jet $p_T$ and pseudo-rapidity $\eta$ spectra, together with the jet mass distributions, have been calculated. The calculated jet observables can probe saturation effects in an IR- and collinear-safe way, with reduced sensitivity to the fragmentation process.

In forthcoming work, we will extend this model to include NLO corrections within the hybrid formalism, a detailed treatment of multiparton interaction (MPI) processes, and a generalization to pA and AA collisions. It would also be an important direction to extend event generators to implement the fluctuation of the saturation scale, as well as two-particle correlations and heavy-flavor probes based on the CGC framework, and to investigate the resulting particle correlation observables, which we leave for future work.\\

\section*{Acknowledgment}
We gratefully acknowledge T.~Chujo for fruitful discussions.
The work was supported by JSPS KAKENHI Grant No.~JP24H00003.

\bibliographystyle{apsrev4-2}
\bibliography{ref}
\end{document}